\documentclass[aps,prb,twocolumn,showpacs]{revtex4-1}
\usepackage{amsmath,amssymb,graphicx}

\usepackage[T1]{fontenc}
\usepackage{chancery}

\usepackage{xcolor}
\usepackage{physics}
\usepackage[bottom]{footmisc}

\IfFileExists{newtxtext.sty}   {\usepackage{newtxtext,newtxmath}}
{\IfFileExists{stix.sty}
	{\usepackage{stix}}
	{\IfFileExists{mathptmx.sty}
		{\usepackage{mathptmx}}{} } }

\usepackage{textcomp}

\usepackage{bm}

\IfFileExists{siunitx.sty}{\usepackage{booktabs,siunitx}}{}

\usepackage{color}
\definecolor{LinkColor}{rgb}{0.256,0.439,0.588}
\usepackage{hyperref}
\hypersetup{
	pdfauthor={Tsung-Cheng Lu},
	pdftitle={Renyi Entropy of Chaotic Eigenstates},
	colorlinks=true,
	citecolor=LinkColor,
	linkcolor=LinkColor,
	urlcolor=LinkColor
}

\usepackage{pifont}
\usepackage{epstopdf}
\usepackage{caption}
\usepackage{subcaption}
\renewcommand{\raggedright}{\leftskip=0pt \rightskip=0pt plus 0cm}
\let\vec\mathbf

\newcommand{\be}{\begin{equation}}
\newcommand{\ee}{\end{equation}}                  
\newcommand{\bea}{\begin{eqnarray}}
\newcommand{\eea}{\end{eqnarray}}
\newcommand{\beas}{\begin{eqnarray*}}
	\newcommand{\eeas}{\end{eqnarray*}}
\begin{document}
	
    \title{Structure of Quantum Entanglement at a Finite Temperature Critical Point}

	\author{Tsung-Cheng Lu}

    \author{Tarun Grover {\vspace{0.3cm}}}

\affiliation{Department of Physics, University of California at San
		Diego, La Jolla, CA 92093, USA}
	\begin{abstract}
		\noindent
We propose a scheme to characterize long-range quantum entanglement close to a finite temperature critical point using tripartite entanglement negativity. As an application, we study a model with mean-field Ising critical exponents and find that the tripartite negativity does not exhibit any singularity in the thermodynamic limit across the transition. This indicates that the long-distance critical fluctuations are completely classical, allowing one to define a `quantum correlation length' that remains finite at the transition despite a divergent physical correlation length. Motivated by our model, we also study mixed state entanglement in tight-binding models of bosons with  $U(1)$ and time-reversal symmetry. By employing Glauber-Sudarshan `P-representation', we find a surprising result that such states have zero entanglement.
%The entanglement of quantum models at finite temperature can be characterized via mixed state entanglement measures such as  negativity. If these models host a finite temperature transition, one may ask whether there is long-range entanglement at the finite temperature critical point? and implement it for a model which exhibits a phase transition with mean-field theory critical exponents. We find that despite the singular nature of the short-distance component of negativity, there is no singularity in the long-range part indicating that the universal, long-distance part of critical fluctuations is completely classical. Motivated by our model, we also study mixed state entanglement of tight-binding models of bosons, and find a surprising result that such states are separable if the system possesses $U(1)$ and time-reversal symmetry. 
	\end{abstract}
	
%	\pacs{71.10.-w,71.10.Hf,75.40.Cx,75.40.Mg}
		\maketitle
	
%	\tableofcontents

\section{Introduction} \label{sec:intro}

%The concept of \textit{universality} is central to quantum mechanics of many-body systems. Although one can write down an infinite number of distinct physical Hamiltonians, all models belonging to the same phase of matter share certain `universal' properties which leads to enormous simplification in organizing our knowledge. An attractive platform to explore universality is phase transitions across two different phases of matter whereby the corresponding `critical exponents', which characterize scaling behavior of various operators, depend only on the symmetry and the space-time dimension of physical models. 

Qualitatively, there are two distinct classes of phase transitions in quantum mechanical Hamiltonians: those that occur at the absolute zero temperature, and those at a finite (i.e. non-zero) temperature. Heuristically, the zero temperature phase transitions result due to quantum fluctuations while the finite temperature ones typically result from thermal fluctuations. For example, consider the transverse field Ising model on a $d$-dimensional hypercubic lattice, $H_{\textrm{TFI}} = -\sum_{<i,j>} Z_i Z_j  - h \sum_i X_i$. This Hamiltonian supports two phases: a ferromagnetic phase and a paramagnetic phase. The critical exponents associated with the zero temperature  transition belong to the $d+1$-dimensional Ising universality while those for the finite temperature  transition belong to the $d$-dimensional Ising universality i.e. at finite temperature, one may as well set $ h= 0$ to obtain the critical exponents \cite{suzuki1976,sachdev2011quantum}. This is consistent with the conventional wisdom that quantum mechanics does not play any role in the long-distance equilibrium physics of finite temperature phase transitions. 

However, there also exist models such as the four-dimensional toric code \cite{dennis2002} which host finite temperature `quantum memory', that is, one can encode a qubit non-locally in this model at finite temperature such that it is well protected for an infinite time even when coupled to a heat bath. This model also exhibits a finite temperature phase transition across which the quantum memory is destroyed. Therefore, one  suspects that in this model, both at and below the critical temperature there exist intrinsically quantum effects even at long-distances.  This raises the question: is there a quantity that sharply distinguishes the finite temperature transition in a transverse field Ising model from that in the 4D Toric code?  

A related question is: how easy is it to prepare thermal (mixed) states on a classical computer? It has been argued that a thermal state can be prepared efficiently \textit{if} the system does not possess a finite temperature quantum memory, \textit{and} has a short correlation length \cite{swingle2016_1, swingle2016_2, Brandao2019}. Similar results have been argued for the preparation of thermofield double state which corresponds to a purification of a thermal density matrix \cite{wu2018variational,martyn2018, Cottrell2019, maldacena2018eternal,chapman2019complexity}. However, if the long-range correlations in a system arise purely due to classical effects (e.g., consider $H_{\textrm{TFI}}$ at the finite temperature critical point for $|h| \ll 1$), one might wonder if the corresponding state can again be  prepared efficiently?
% contrary to the aforementioned intuition that perhaps quantum mechanics is irrelevant for universal aspects of  finite temperature phase transitions.

%In this work, we  primarily utilize  a measure of mixed-state entanglement called entanglement negativity \cite{eisert99, vidal2002} (henceforth just `negativity' for brevity).

The above discussion motivates us to ask: How does one separate quantum-mechanical correlations from classical correlations at a finite temperature, and in particular, in the vicinity of a phase transition? At zero temperature, a system can typically be described by a pure state and correspondingly, the von Neumann entropy of a reduced density matrix corresponding to a subsystem is a faithful measure of long-range entanglement in the critical ground state \cite{Holzhey94, Calabrese04, Ryu_2006, casini07, metlitski2009}. In contrast, at a finite temperature $T$, the system is described by a thermal (i.e. Gibbs) state $\propto e^{- H/T}$, which is a mixed density matrix.  To probe an intrinsic quantum correlation at finite temperature, one must therefore resort to an entanglement measure for mixed states \cite{werner1989,horodecki_revmodphys}. To this end, here we will employ `entanglement  negativity' (henceforth just `negativity' for brevity) which has the property that it is an entanglement monotone and unlike most other mixed-state measures, does not require optimizing a function over all possible quantum states \cite{eisert99, vidal2002, plenio2005, horodecki_revmodphys}. As shown in Ref. \cite{sherman2016}, one very interesting property of negativity is that for thermal states of local Hamiltonians it satisfies an `area-law', akin to the von Neumann entanglement entropy of (pure) ground states of gapped Hamiltonians \cite{bombelli1986, srednicki1993_area, eisert_rmp_arealaw}, and in strong contrast to the volume law for pure finite energy density eigenstates \cite{deutsch1991, srednicki1994chaos, rigol2008, rigol2012, deutsch2010, rigol_review, garrison2015does}.

A recent work, Ref.\cite{lu2018singularity}, constructed a class of exactly solvable models which host finite-$T$ order-disorder transition and for which negativity can be calculated analytically. It was found that in all models considered, whenever negativity is non-zero in the vicinity of the transition, it is a singular function of the tuning parameter driving the transition, despite the fact that these phase transitions are driven purely by thermal fluctuations and do not host finite-$T$ quantum memory. As argued in Ref.\cite{lu2018singularity}, the area law coefficient of negativity receives contribution from the expectation value of local operators close to the entangling boundary, and since the expectation value of local operators is singular across the transition, the area-law coefficient is singular as well. 
%area law is sensitive to short-distance quantum correlations close to the bipartition boundary. Since even expectation of  local operators, such as energy density, are singular across the transition, one finds that the area-law coefficient of the negativity will also be singular generically. \textcolor{red}{Define LA/ell}% It is also at variance with expectations from Ref.\cite{sherman2016}  where numerical calculations on finite sized systems for the 2+1-dim quantum Ising model suggested that negativity is analytic across a critical temperature. 
\begin{figure}
	\centering
	\includegraphics[width=0.35\textwidth]{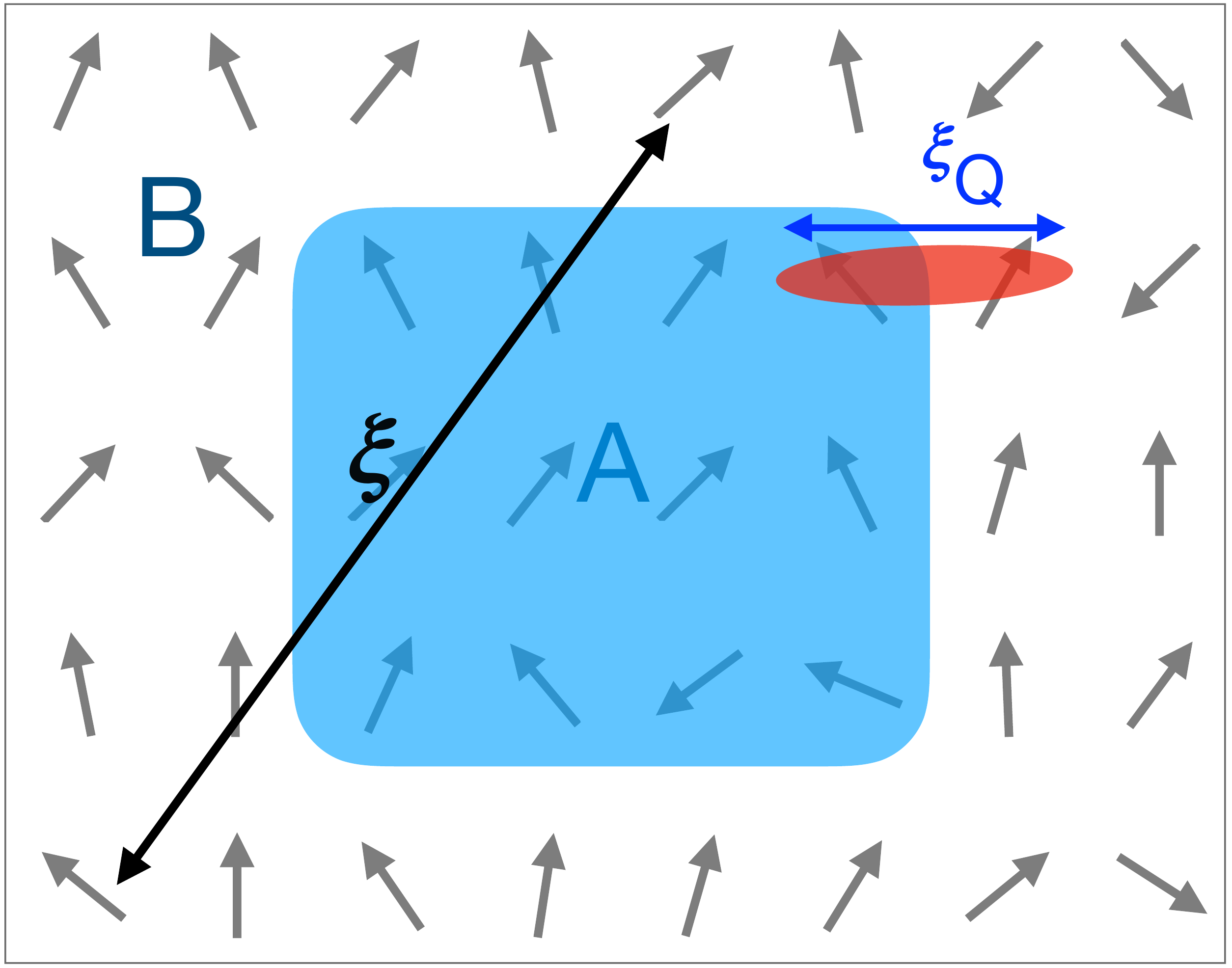}
	\caption{At a finite temperature $T$, the physical correlation length $\xi$ will generically be different from the length scale  $\xi_Q$ over which quantum correlations exist (denoted by red blob). At a finite-$T$ critical point, $\xi$ diverges while $\xi_Q$ can continue to remain finite. As discussed in the main text, mixed-state entanglement  between subregions $A$ and $B$ can in principle distinguish $\xi_Q$ from $\xi$.}
	\label{fig:lengths}
\end{figure}

The aforementioned singularity in area-law coefficient across a finite-$T$ transition leaves open the question of extracting purely quantum correlations that are \textit{sensitive only to long-distance physics}, unlike the area-law coefficient of negativity which certainly depends on short-distance physics despite containing information about critical exponents. This is the topic of this paper. Inspired by the methods used to extract universal entanglement encoded in the ground states of gapped Hamiltonians  \cite{Kitaev06_1, levin2006detecting},  we propose a tripartite negativity to probe long-distance, universal quantum correlations at finite temperature. We study the tripartite negativity, denoted as $\Delta_3 E_N$ below, for a simple model that  exhibits singularity in correlation functions, von Neumann entropy as well as the area-law coefficient of negativity across a finite-$T$ transition. We find that $\Delta_3 E_N$ \textit{completely  cancels out the aforementioned singularity associated with the transition}, and is exponentially small in the system size $\Delta_3 E_N \sim e^{-L/\xi_Q}$ where $\xi_Q$ defines a `quantum correlation length' which, in contrast to the physical correlation length, does not diverge at the finite-$T$ transition (Fig.\ref{fig:lengths}). As $T \rightarrow 0$, $\xi_Q $ diverges resulting in a non-zero $\Delta_3 E_N$ which corresponds to the universal non-zero subleading term for Renyi entropy $S_{1/2}$ at the quantum phase transition. Note that at $T = 0$, for a gapped, topological ordered phase $\Delta_3 E_N$ also equals the topological entanglement entropy  \cite{vidal2013, Castelnovo2013}.

%Or in other words, although classical correlations are gapless at the finite temperature critical point, the quantum correlations continue to remain gapped.
%This model in fact corresponds to the mean-field approximation with Gaussian fluctuations to the finite temperature Ising critical point for the quantum Ising model. 
Partly inspired by our model, we also study mixed state entanglement in tight-binding models of free bosons with time-reversal and $U(1)$ symmetry. We show that the corresponding thermal state is separable, and therefore, \textit{any} measure of mixed state entanglement for such a state, such as entanglement of negativity or entanglement of formation, is zero. 

The paper is organized as follows: In Sec.~\ref{sec:general}, we discuss the general structure of negativity for local Hamiltonians. In Sec.~\ref{sec:critical}, we introduce our scheme for calculating the universal part of negativity, and implement it for a model that shows a finite temperature transition. In Sec.~\ref{sec:zeroTcrossover} we discuss cross-over towards the zero temperature quantum phase transition, and in Sec.\ref{sec:eigenspectrum} we discuss the eigenvalues and eigenfunctions of correlation matrix that determines negativity.  In Sec.~\ref{sec:bosonic_tb}, we discuss our aforementioned result on separability of bosonic Gaussian states with $U(1)$ and time-reversal symmetry. In Sec.~\ref{sec:discuss}, we conclude with a summary and possible implications of our results.

%We find that the negativity of a Gaussian state of bosons with time-reversal and $U(1)$ symmetry is identically zero. In fact, we prove a much stronger result: 

% To be specific, consider three subregions $A$, $B$, and $C$ as shown in Fig.\ref{fig:partition}, we define the universal part of the negativity as $E_{N,\textrm{KP}} = E_{N,A} + E_{N,B}  + E_{N,C}  - E_{N,AB}  - E_{N,BC}  - E_{N,CA}  + E_{N,ABC}$, where $E_{N,A}$ denotes the negativity between the region $A$ and its complement, and similarly for all the other quantities present in $E_{N,\textrm{KP}}$. Contrary to the observation that negativity is singular across finite temperature phase transitions whose universalities are determined solely by classical physics, we expect $E_{N,\textrm{KP}}$ is not singular across such transitions.  

%We study two models in this work to demonstrate the advantage of the subtracted negativity. The first model exhibits the Ising phase transition at finite temperature within a gaussian approximation. In this model, we find that while the area-law part of the negativity is singular across a critical temperature, the universal component of negativity $E_{N,\textrm{KP}}$ decays exponentially with system size. In strong contrast, the von Neumann entropy after employing the subtraction scheme exhibits a logarithmic singularity across the phase transition, consistent with our intuition that the singularity in von Neumann entropy as well as physical correlations (such as the correlation length) is \textit{solely} due to classical correlations. 

\section{Negativity across finite-T critical points} \label{sec:finiteT_ngvt}

\subsection{General structure of negativity: local versus non-local contributions} \label{sec:general}

Given a state $\rho$ acting on the Hilbert space $\mathcal{H}_A\otimes\mathcal{H}_{\bar{A}}$, the negativity is defined as $E_N(A)= \log \norm{\rho^{T_A}}_1$.   Due to the area-law of negativity for thermal states of local Hamiltonians (Ref.\cite{sherman2016}), the problem to characterize the universal part of their negativity is somewhat analogous to the characterization of long-distance entanglement in ground states of gapped Hamiltonians. Following Ref.\cite{turner2011}, we consider a coarse-grained, continuum description, and write 
\be 
E_N = E_{N,\textrm{local}} + E_{N, \textrm{non-local}},
\ee
where $E_{N,\textrm{local}}$ is expressible as a sum of local terms along the entangling surface:  $E_{N,\textrm{local}} = \int_{\partial A} F(\{\kappa, \partial_i \kappa,...,\})$ where $\kappa$  is the local curvature along the entangling surface. Similar to the von Neumann entropy of pure states, negativity of mixed states satisfies $E_N(A) = E_N (\bar{A})$. This is because
\begin{equation}
E_N(A)= \log \norm{\rho^{T_A}}_1=\log   \norm{ (\rho^{T_{\bar{A}}})^{T}}_1=\log   \norm{ \rho^{T_{\bar{A}}}}_1, 
\end{equation}
where the last equality results from invariance of eigenspectrum under matrix transpose operation.  Since under the exchange $A \leftrightarrow \bar{A}$, the curvature $\kappa \leftrightarrow - \kappa$, $F$ must be an even functional of the curvature $\kappa$, and consequently $E_{N,\textrm{local}} = \alpha_{d-1} L_A^{d-1}  + \alpha_{d-3} L_A^{d-3} + ...$, i.e., only alternate terms in the expansion in terms of $L_A$ are allowed, again similar to the discussion of gapped ground states (Ref.\cite{turner2011}). One implication of this is that in two-dimensions, a non-zero constant term $\gamma$ in $E_N \sim L_A - \gamma$ necessarily implies a non-zero $E_{N,\textrm{non-local}}$. 

Again motivated by the theory of gapped ground states \cite{Kitaev06_1, levin2006detecting}, below we use a subtraction scheme to cancel out  $E_{N, \textrm{local}}$ to understand the behavior of $E_{N,\textrm{non-local}}$, and specifically whether it has a non-zero value at a finite-$T$ phase transition in the thermodynamic limit. It's worth emphasizing that the coefficients $\alpha_i$ that enter $E_{N,\textrm{local}} $ will generically be singular functions of the tuning parameter driving the transition. This is because these coefficients will depend on the expectation value of local operators, such as energy density, which themselves are a singular function of the tuning parameter. This leads to the singularity in the area-law coefficient of negativity as discussed in Ref.\cite{lu2018singularity}, and which we will again encounter below. 

It is also important to note that  for pure states which exhibit power-law correlations, such as a ground state corresponding to a conformal field theory, or a Fermi surface, the non-local part of entanglement is necessarily non-zero \cite{callan94,Holzhey94, Calabrese04, Gioev06, wolf06, Swingle10}. In contrast, for thermal (mixed) states of quantum systems that display power-law correlations, such as those corresponding to a finite-$T$ phase transition, the entanglement structure close to the transition remains completely unexplored. To that end, we now turn to studying negativity in a specific model that displays a finite-$T$ phase transition.

The aforementioned curvature expansion for negativity relies on a coarse-grained continuum description. For such a description to be valid, one requires that all length scales involved are much larger than the short-distance lattice-cutoff $a$. Close to a finite temperature critical point, the physical correlation length $\xi$ of course satisfies $\xi \gg a$, but as hinted above, we will also encounter a length scale $\xi_Q$ that does not diverge. As we will see below, $\xi_Q \gg a$ at low-temperature, and therefore, the argument is valid over a range of temperatures. This is similar to the discussion in the context of gapped ground states  \cite{turner2011} where the correlation length is assumed to stay large compared to the lattice cut-off.

\subsection{Universal negativity of a model with finite-T phase transition} \label{sec:critical}

We consider a $d$ dimensional cubic lattice of $N=L^d$ sites, where a site at $\vec{r}$ is associated with a degree of freedom described by a canonically conjugate pair $(\phi_{\vec{r}}, \pi_{\vec{r}})$.  The Hamiltonian reads
\begin{equation}\label{eq:H}
H=\frac{1}{2} \sum_{\vec{r}} \left( \pi_{\vec{r}}^2 +m^2 \phi_{\vec{r}}^2 \right) +  \frac{1}{2}  \sum_{\expval{\vec{r},\vec{r}'}}K \left( \phi_{\vec{r}} -\phi_{\vec{r}'}   \right) ^2,
\end{equation}
where the physical mass obeys
\begin{equation}
m(g)=\begin{cases}
\sqrt{g-g_c} \quad \quad ~ \text{for} \quad g>g_c\\
\sqrt{2(g_c-g)}  \quad \text{for} \quad g<g_c .
\end{cases}
\label{eq:mass}
\end{equation}
$g_c$ sets the critical point where correlation length diverges due to the vanishing mass term. The motivation to study this specific model comes from the transverse-field Ising model:  $H_{\textrm{TFI}} = -\sum_{<\vec{r},\vec{r}'>} Z_{\vec{r}} Z_{\vec{r}'}  - h \sum_{\vec{r} } X_{\vec{r} }$. Both $H$ and $H_{TFI}$ have $\mathbb{Z}_2$ symmetry, and $H$ can be thought of as a mean-field approximation to $H_{TFI}$ where  $\pi_{\vec{r}}^2/2 $ plays the role of the transverse field $X_\vec{r}$, and  $\phi_{\vec{r}}$ plays the role of $Z_{\vec{r}}$. The function $m(g)$ corresponds to the effective mass (= inverse correlation length) within the mean-field theory. Close to $g_c$, at finite temperature, the system exhibits long-range correlation functions. For example, in 2d, at $g_c$ $\langle \phi_{\vec{r}}\, \phi_{\vec{r}'} \rangle \sim \log(|\vec{r} - \vec{r}'|)$. Another signature of the divergent correlation length is that the von Neumann entropy $S = - \tr \left( \rho \log \rho\right)$ shows a logarithmic divergence as approaching to $g_c$: $S \sim -\log(m L)$\cite{gelfand1988} as shown in Fig.\ref{fig:von_Neumann} ( see Appendix \ref{sec:entropy_gaussian} for derivation, and footnote \cite{footnote_bc} regarding how the massless limit is taken). Note that, instead of tuning $g$, one can also tune the temperature to drive the finite-$T$ transition (see Fig.\ref{fig:phase_diagram}). All of our results below are unchanged if one simply replaces $g -g_c$ by $(T-T_c)/T_c$.

\begin{figure}
	\centering
	\begin{subfigure}[b]{0.4\textwidth}
				\caption{}
		\includegraphics[width=\textwidth]{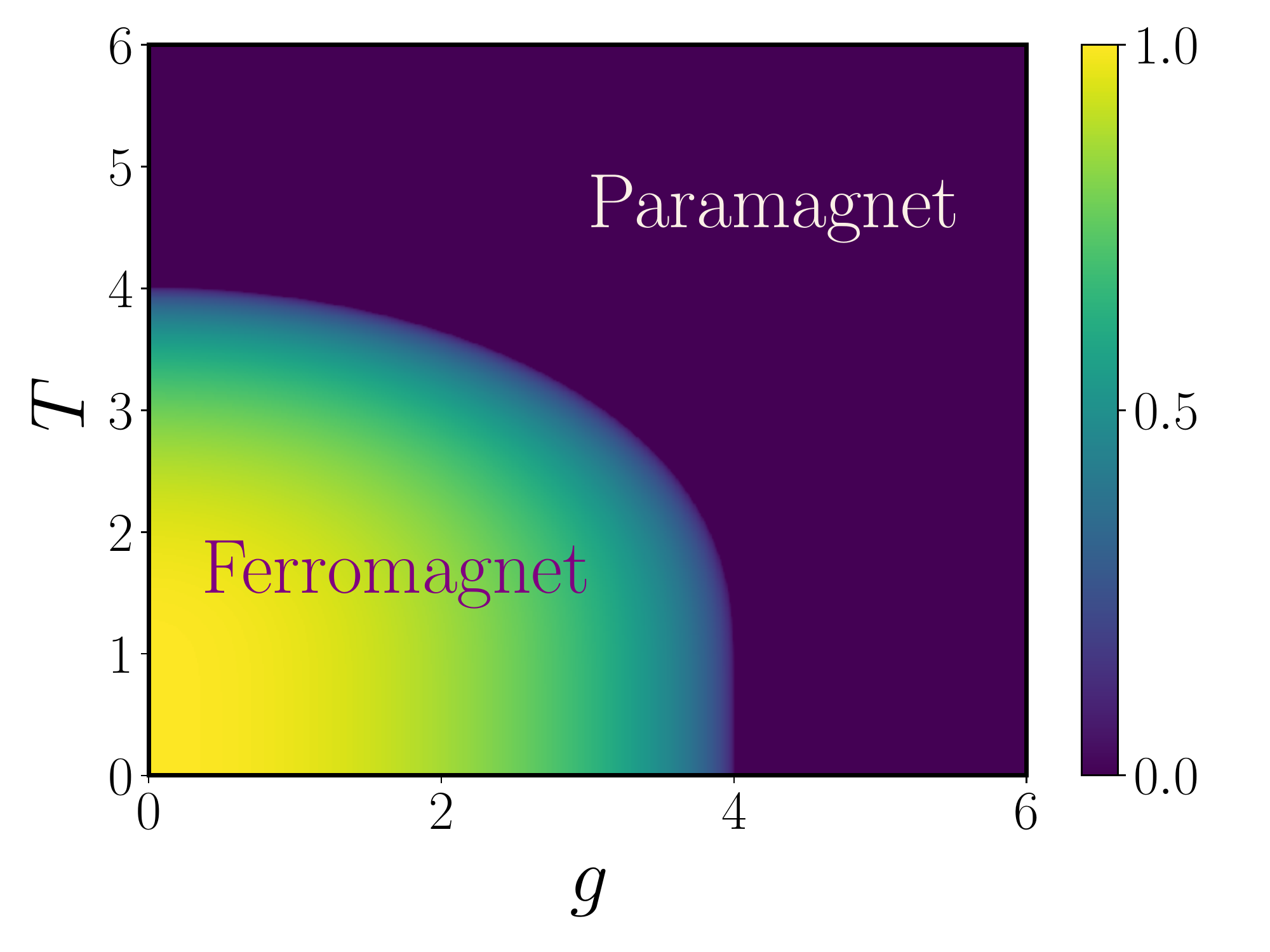}
		\label{fig:phase_diagram}
	\end{subfigure}
\begin{subfigure}[b]{0.4\textwidth}
			\caption{}
\includegraphics[width=\textwidth]{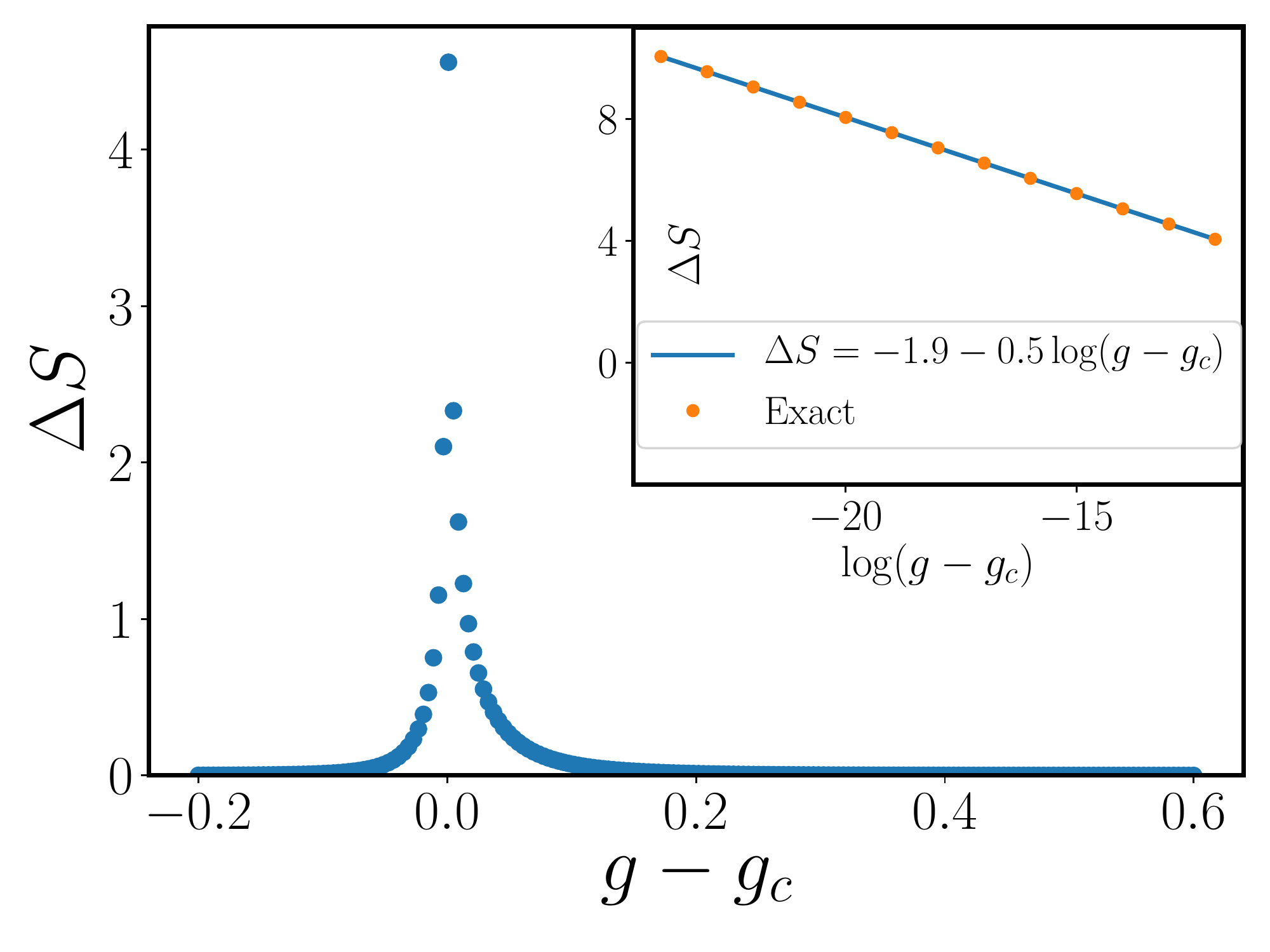}
\label{fig:von_Neumann}
\end{subfigure}
\caption{(a) Phase diagram of the model described by Hamiltonian in Eq.\ref{eq:H}. The color codes the magnitude of the order parameter. (b) Divergence of the universal part of the von Neumann entropy at the finite temperature phase transition obtained by defining $\Delta S = \left[4\,S(L)-S(2L)\right]/3$ to cancel the volume law component.}
%\begin{figure}
%	\centering
%	\includegraphics[width=0.5\textwidth]{von_Neumann.pdf}
%	\caption{Divergence of the universal part of the von Neumann entropy at the finite temperature phase transition.}
%	\label{fig:von_Neumann}
\end{figure}

\begin{figure*}[ht]
	\centering
	\begin{subfigure}[b]{0.33\textwidth}
		\caption{}
		\includegraphics[width=\textwidth]{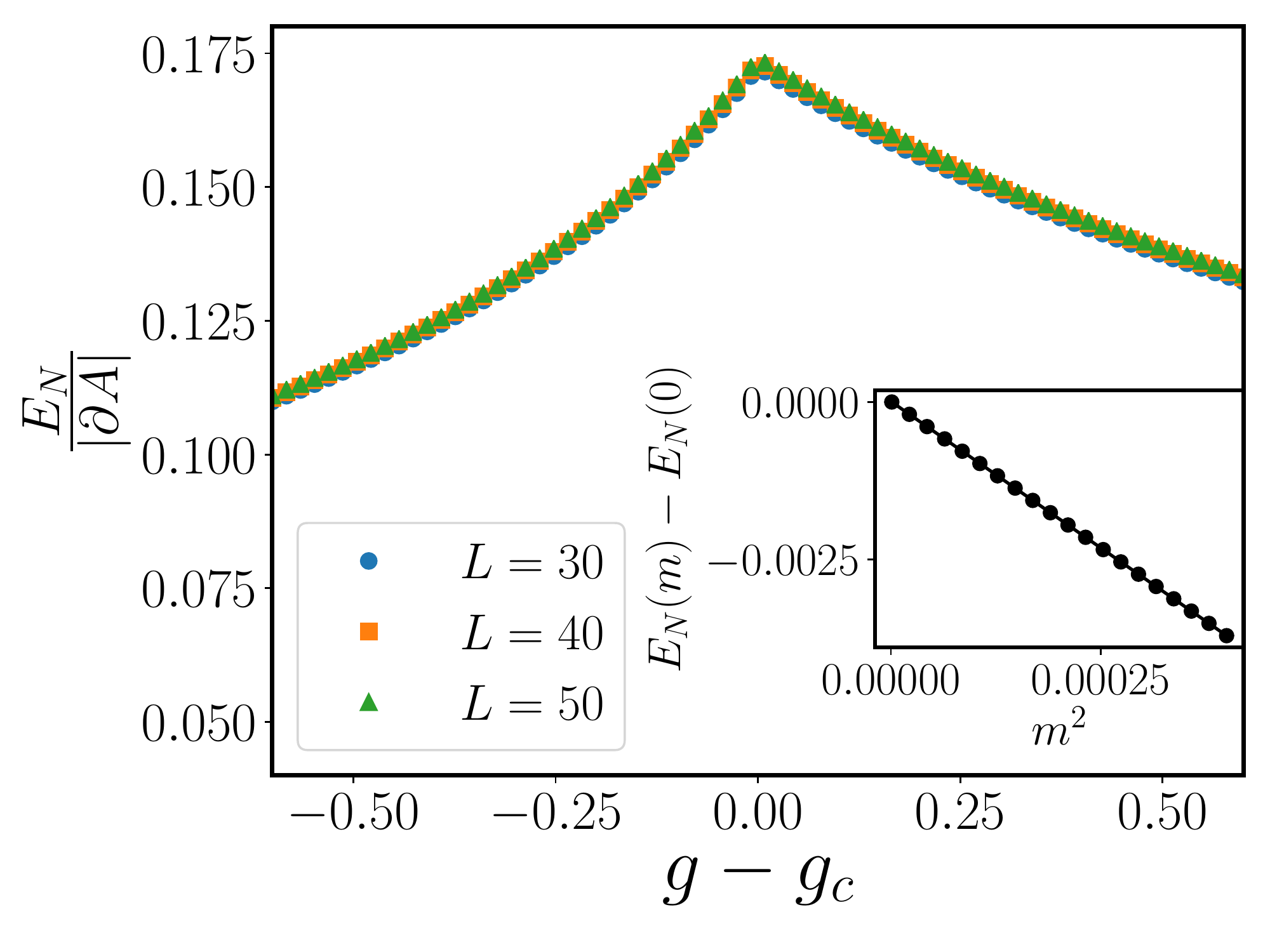}
		\label{fig:nega_area_law}
	\end{subfigure}
	\begin{subfigure}[b]{0.33\textwidth}
		\caption{}
		\includegraphics[width=\textwidth]{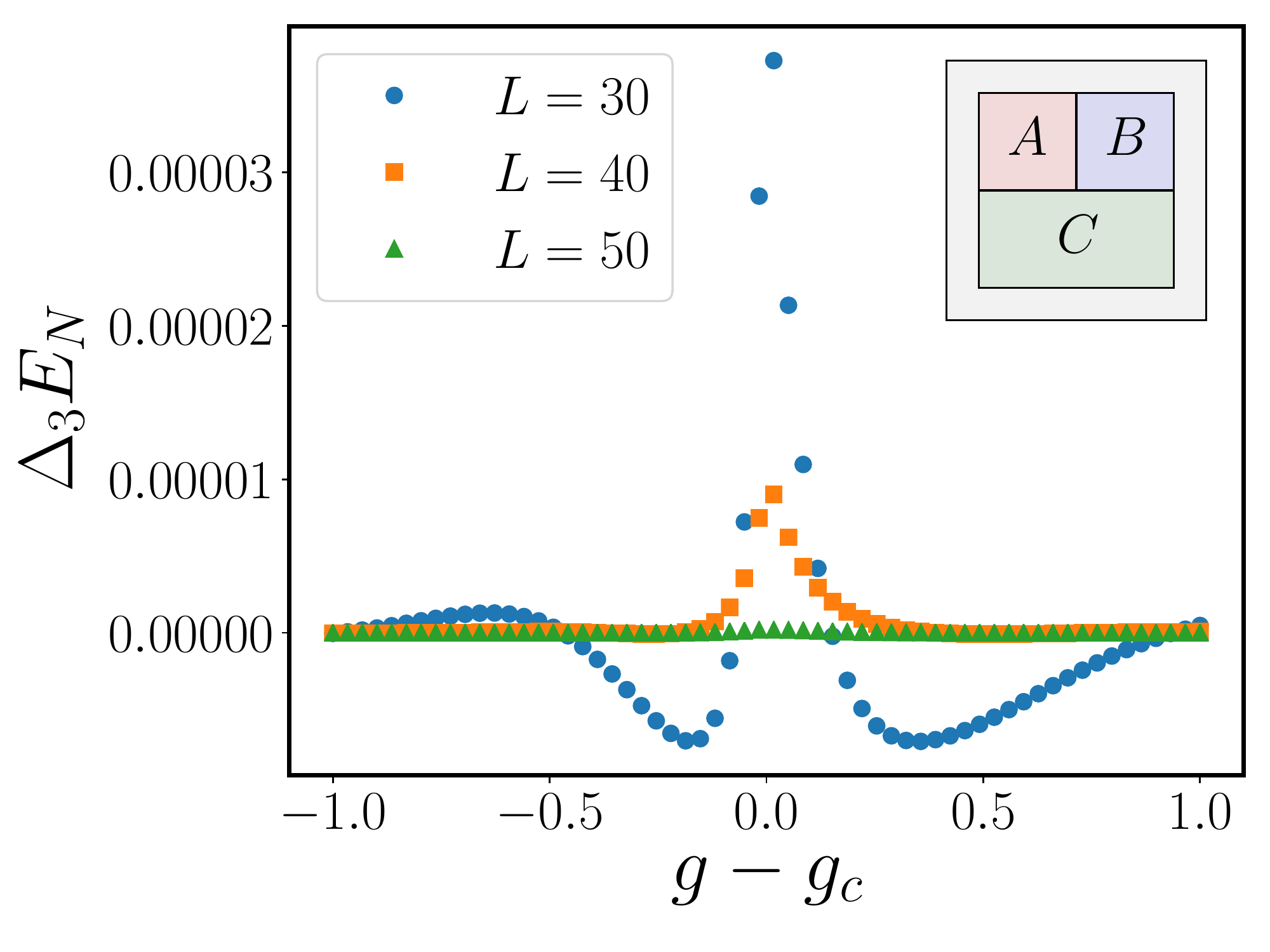}
		\label{fig:nega_kp}
	\end{subfigure}
	\begin{subfigure}[b]{0.33\textwidth}
		\caption{}
		\includegraphics[width=\textwidth]{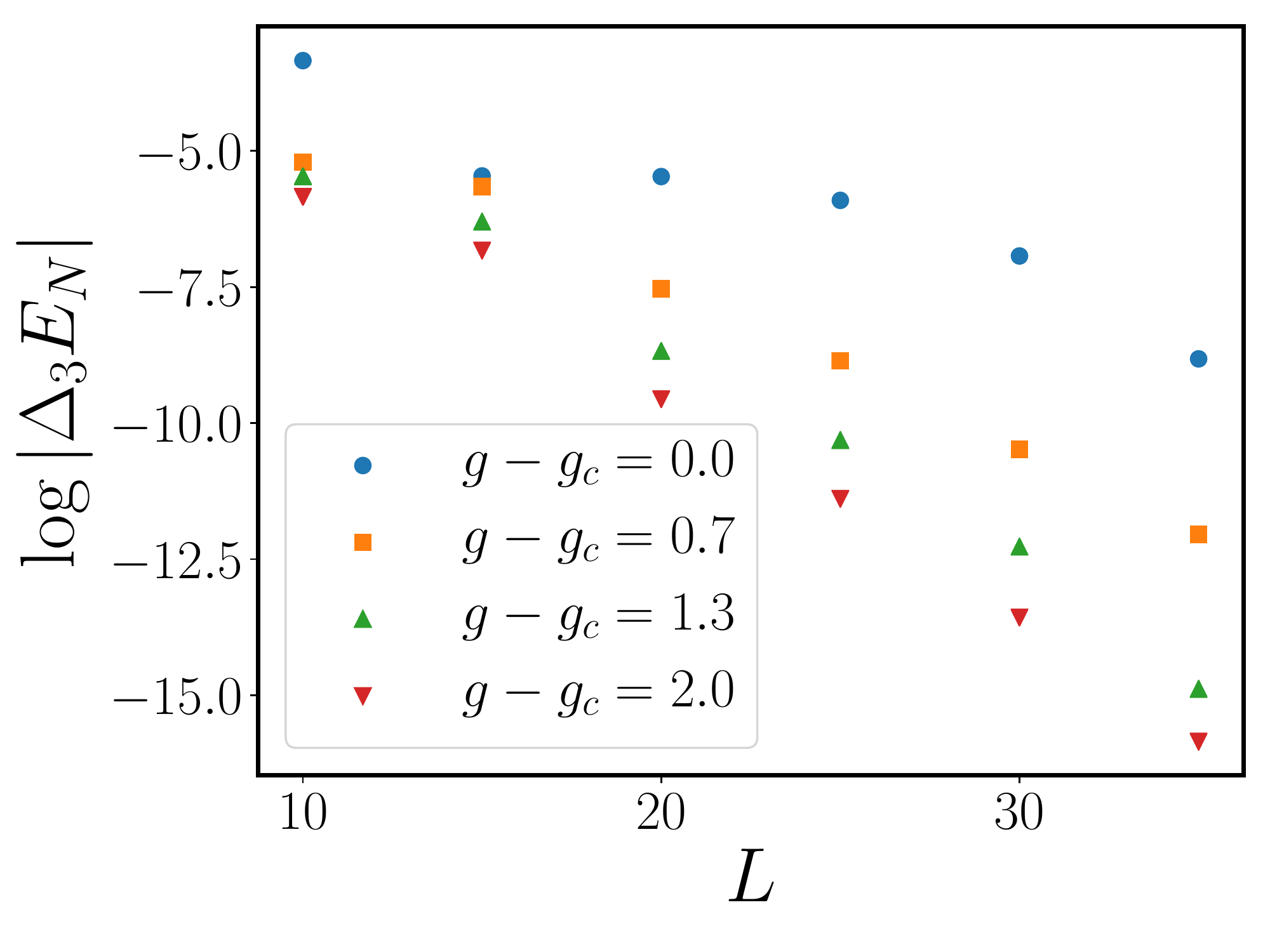}
		\label{fig:nega_kp_scaling}
	\end{subfigure}
	\caption{(a) Area law coefficient of negativity as a function of $g-g_c$ at $T=0.2$. Inset: mass dependence of negativity close to the transition at $T=0.2$. (b) Long-distance component $\Delta_3 E_{N}$ of negativity at $T=0.2$ (c) Finite size scaling of $\Delta_3 E_{N}$ at T=0.1.} 
\end{figure*}

Next, we analyze the structure of quantum correlations close to the transition using entanglement negativity. Since $H$ is quadratic, a thermal state $\rho$ at inverse temperature $\beta$, i.e. $\rho\sim e^{-\beta H}$, is a Gaussian state. This allows an efficient calculation of negativity \cite{audenaert2002entanglement, Ferraro2008, anders2008,Cavalcanti2008, Marcovitch2009} because one can perform partial transposition on $\rho$ using the covariance matrix technique. For concreteness, we consider a two dimensional lattice ($d=2$), and set $K=1$. We first plot the negativity of a subregion $A$ and its complement as a function of $ (g-g_c)$ in Fig.\ref{fig:nega_area_law}. In agreement with the result of Ref.\cite{lu2018singularity}, we find that the area law coefficient of the negativity is singular at $g=g_c$. As mentioned above, this singularity originates from the singular behavior of the expectation value of local operators close to the entangling boundary \cite{lu2018singularity}. What is the precise nature of this singularity? We numerically find that in the limit $a<<\beta << 1/m$, at the lowest order, the singular part of area-law coefficient is proportional to  $m^2$. Since $m^2 \sim |t|$ where $t$ is the thermal tuning parameter (either $T-T_c$ or $g-g_c$),  
\begin{equation}
E_{N, \textrm{local}}= \left[ \alpha+ b_{\pm} |t|  \right]L^{d-1}
\end{equation}
where $\alpha$ is an analytic function of underlying parameters, and $b_{+} ~ (b_{-})$ is coefficient of $|t|$ for $g > g_c$ ($g < g_c$) with $b_{-}/b_{+}  = 2$ due to the singular dependence of mass close to the transition (Eq. \ref{eq:mass}). As shown in Appendix.\ref{sec:two_sites}, one can gain some intuition for the $m^2$ dependence by analytically studying the negativity between two sites. We comment on the relation of this singularity to critical exponents in Sec.\ref{sec:discuss}.

To isolate  long-distance quantum correlations, we now define a quantity analogous to `topological entanglement entropy' in the context of ground-states of gapped Hamiltonians \cite{Kitaev06_1, levin2006detecting} that cancels out the $E_{N,\textrm{local}}$  component of the negativity. Consider three subregions $A,B$ and $C$, and define a tripartite negativity,
\begin{equation}
\Delta_3 E_{N} = E_{N,A} + E_{N,B}  + E_{N,C}  - E_{N,AB}  - E_{N,BC}  - E_{N,CA}  + E_{N,ABC} \label{eq:E3N}
\end{equation}
Here $E_{N,A}$ denotes the negativity between the region $A$ and its complement, and similarly for all the other quantities present in $\Delta_3 E_{N}$. On a square lattice of size $L$ with the periodic boundary condition, we consider the partition  shown in the inset of Fig.~\ref{fig:nega_kp}, where $A,B$ and $A\bigcup B \bigcup C$ are squares of size $2/5L,2/5L$ and $4/5L$ respectively. Fig.~\ref{fig:nega_kp} shows the dependence of  $\Delta_3 E_{N}$ on $g-g_c$ for various system sizes. Despite the fact that each of the seven individual terms that enters the definition of $\Delta_3 E_{N}$ (Eq.\ref{eq:E3N}) is singular (see Fig.\ref{fig:nega_area_law}),   one finds that $\Delta_3 E_{N}$ itself does not exhibit any singularity across $g_c$, upto terms that are exponentially small in the total system size $L$. In fact, $\Delta_3 E_{N}$ itself vanishes exponentially with $L$ at all non-zero temperatures (see Fig.\ref{fig:nega_kp_scaling}): 

\be 
\Delta_3 E_N \sim e^{-L/\xi_Q} 
\ee 
which defines a  `quantum correlation length' $\xi_Q$ that remains finite even at the critical point. The peak in $\Delta_3 E_N $ at the critical point (Fig.\ref{fig:nega_kp}) indicates that $\xi_Q$ is largest at the critical point. We discuss the detailed behavior of $\xi_Q$ below for the case of a straight bipartition without any corners.  It is worth emphasizing that the singularity in quantities that are sensitive both to classical and quantum correlations survives in the thermodynamic limit after an analogous subtraction scheme. For example, an analogously defined tripartite von Neumann entropy $\Delta_3 S$ continues to show singular behavior identical to Fig.\ref{fig:von_Neumann}.

\begin{figure*}[ht]
	\centering
	\begin{subfigure}[b]{0.33\textwidth}
		\caption{}
		\includegraphics[width=\textwidth]{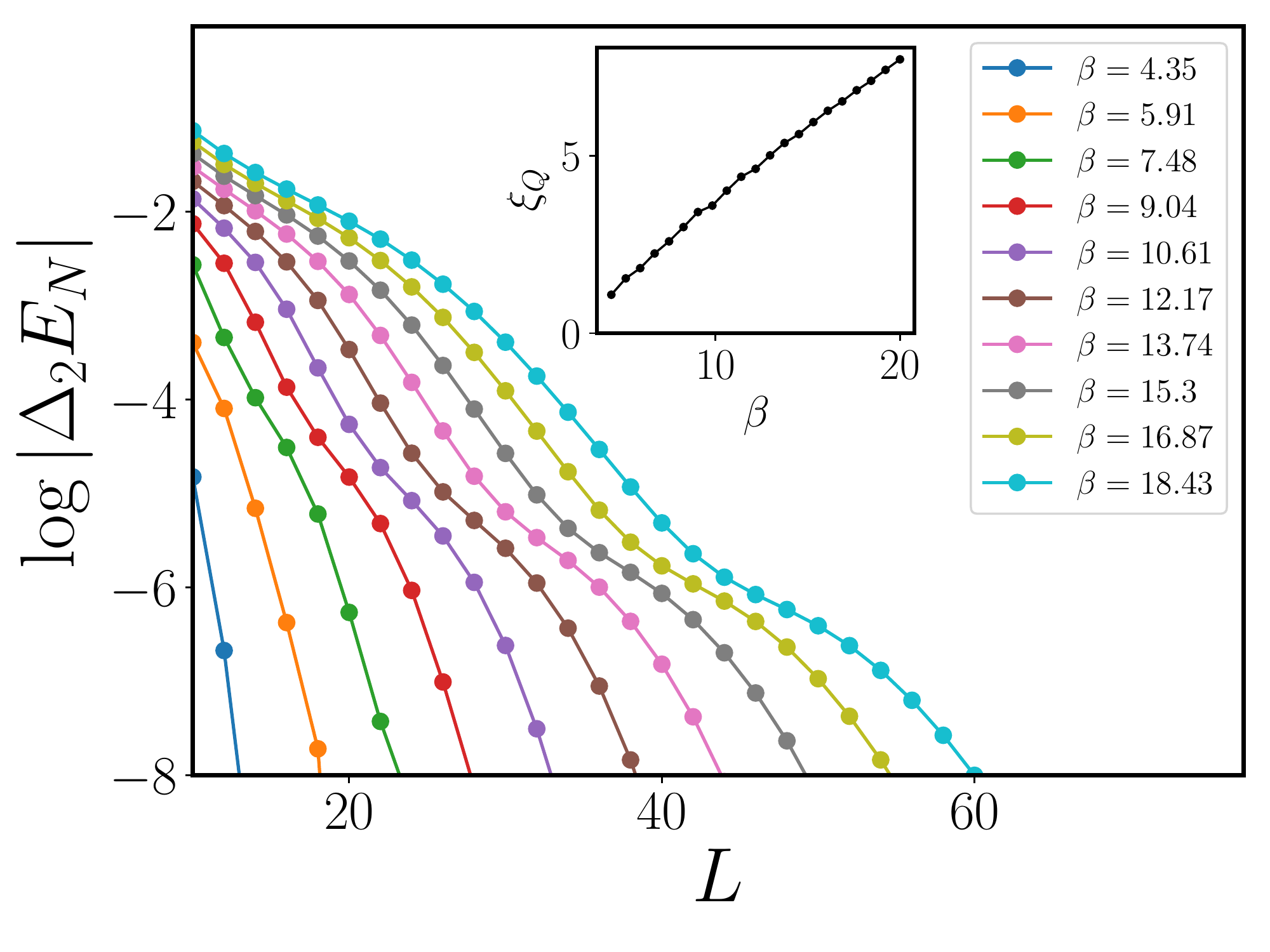}
		\label{fig:flat_subtracted}
	\end{subfigure}
	\begin{subfigure}[b]{0.33\textwidth}
		\caption{}
		\includegraphics[width=\textwidth]{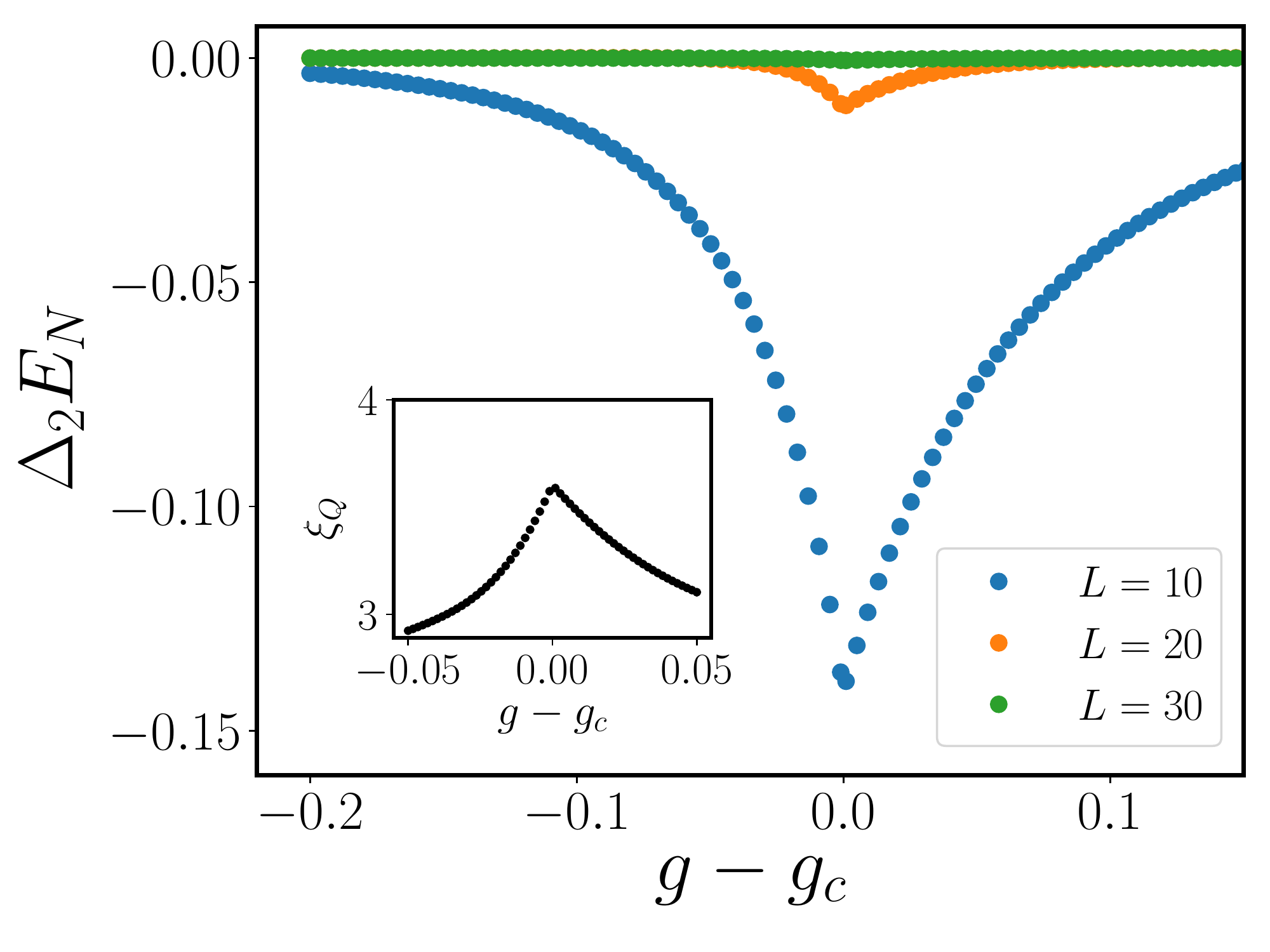}
		\label{fig:flat_subtracted_xi}
	\end{subfigure}
	\begin{subfigure}[b]{0.33\textwidth}
		\caption{}
		\includegraphics[width=\textwidth]{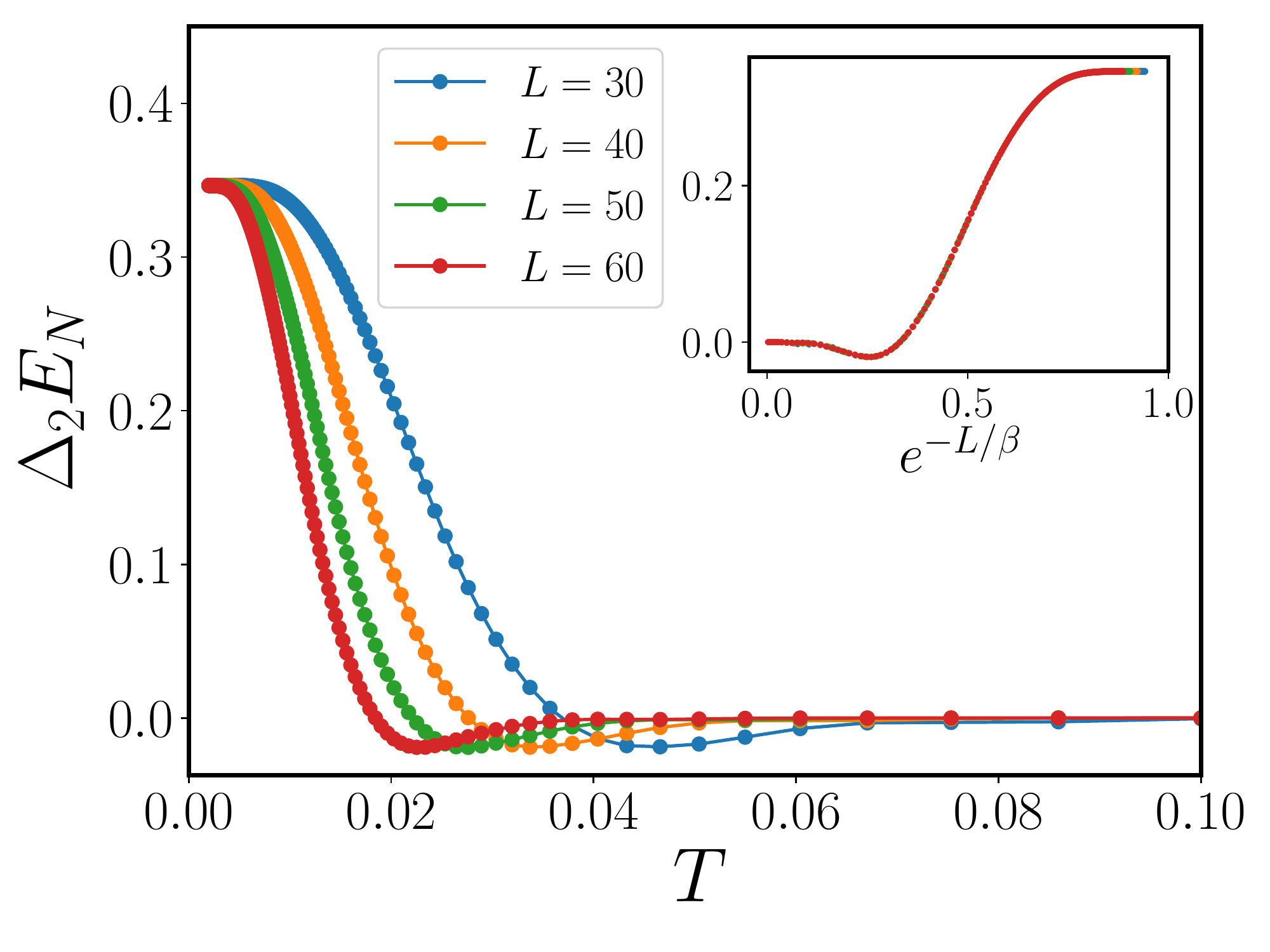}
		\label{fig:cross_over}
	\end{subfigure}
	\caption{(a) The scaling of the subtracted negativity $\Delta_2E_N=E_N (2L) - 2E_N(L)$ as a function of $L$ at the critical point for various inverse temperature $\beta$. Inset: quantum correlation length $\xi_Q$ extracted via the scaling $\Delta_2E_N \sim e^{-L/\xi_Q}$ as a function of $\beta$. (b) Behavior of $\Delta_2E_N$ in the vicinity of the transition. Inset: Length scale $\xi_Q$ at temperature $T=0.1$. (c) Scaling of $\Delta_2E_N$ at the critical point as a function of $T$ for various system sizes. Inset: scaling collapse of the data when plotted as a function of $e^{-L/\beta}$. For this plot, we set $m=0$ and chose anti-periodic boundary condition along the y-direction, and periodic along x-direction. } 
\end{figure*}

A partial analytical understanding of finiteness of $\xi_Q$ at the critical point is provided by considering a bipartition without any curvature or corners by dividing a torus of size $L\times L$ into two strips of equal size $L/2 \times L$. Based on the discussion above, we expect that for such bipartition, $E_{N,\textrm{local}} \propto L^{d-1}_A$, i.e., it is strictly an area-law without any subleading corrections. The universal part of the negativity can now be extracted by studying $\Delta_2 E_N = E_N (2L) - 2E_N(L)$ which cancels out the aforementioned  $E_{N,\textrm{local}}$ contribution, leaving only $E_{N,\textrm{non-local}}$, the term of interest. The subtraction scheme based on $\Delta_2 E_N $ is conceptually quite similar to that based on $\Delta_3 E_N$, the former is more suited towards a bipartition without any curvature, while the latter is more general. Note that setting $m=0$ explicitly leads to numerical instability in the diagonalization of the covariance matrix, which we regularize by setting $m=10^{-5}$, and  confirm that a further decrease of the mass (while keeping it non-zero) does not change the numerical value of negativity. We find that in this massless limit, $\Delta_2 E_N$ continues to decay exponentially with system size $L$ for all finite temperatures (Fig.\ref{fig:flat_subtracted}), similar to the behavior of $\Delta_3 E_N$ studied above. Furthermore, we find that the quantum correlation length scale $\xi_Q$ is roughly proportional to the inverse temperature $\beta$ as shown in the inset. Using conservation of momentum along the $\hat{y}$ direction, one can express the negativity for this bipartition as the sum  of negativities corresponding to the following 1D Hamiltonians $H_{1D, k_y }$ labeled by $k_y$, the momentum along the $\hat{y}$ direction (see Appendix.\ref{sec:mapping1d} for derivation):

\begin{equation}
H_{1D, k_y }= \frac{1}{2} \sum_{x} \left( \pi_{x}^2 +m_{k_y}^2 \phi_{x}^2 \right) +  \frac{1}{2}  \sum_{\expval{x,x'}}K \left( \phi_{x} -\phi_{x'}   \right) ^2, 
\end{equation}
\noindent where $m^2_{k_y} = 4K \sin[2](\frac{1}{2}k_y )$. One expects that any non-local contribution to negativity can arise only when the mass $m_{k_y} = 0$, i.e., the contribution of $k_y = 0$ mode. Curiously, the contribution of $k_y = 0$ mode is identical to negativity corresponding to  the thermal state of a central charge $c = 1$ 1+1-D CFT studied in Ref.\cite{calabrese2015}. Using results from Ref.\cite{calabrese2015}, one finds that 
\begin{equation}
E_{N, k_y=0} =\frac{1}{2} \ln \left[  \frac{\beta}{\pi a} \sinh\left( \frac{\pi L_A }{\beta }\right)   \right]   -  \frac{\pi L_A}{2\beta} + f(e^{-2\pi L_A/ \beta})  +2 \ln c_{1/2},
\end{equation}
where $f$ is a universal scaling function which tends to a constant when $L_A \gg \beta$, $c_{1/2}$ is not universal and $a$ is the lattice constant. When $L_A \gg \beta$, the expression can be written as 
\begin{equation}
=\left[\frac{1}{2}\log \left( \frac{\beta}{2\pi a }  \right) +c_{1/2} \right] + \left[\log\left( 1- e^{-2\pi L_A/\beta }  \right) + f(e^{-2\pi L_A/\beta }) \right].
\end{equation}
This expression implies that  when $L_A/\beta \rightarrow \infty$, $E_{N, k_y=0}$ approaches a constant value over a characteristic length-scale $\beta$ in line with our numerical results for $\Delta_2 E_N$.

We also study the behavior of $\xi_Q$ in the vicinity of the transition, and find that it is maximum at the transition and exhibits a cusp singularity (Fig.\ref{fig:flat_subtracted_xi}). This is expected since $\xi_Q$ is a function of the mass $m$ which itself is singular across the transition.

\subsection{Approach to Quantum Critical Point} \label{sec:zeroTcrossover}

The quantum correlation length diverges as $T \rightarrow 0$, and in the (pure) ground state the negativity equals $S_{1/2}$ where $S_n = - \frac{1}{n-1}\log \trace \rho_A^n$ is the $n$'th Renyi entropy. Since a massless scalar has long-range entanglement in its ground state which is reflected in Renyi entropies as well, one expects that the non-local part of negativity will be non-zero in the ground state. Fig.\ref{fig:cross_over} shows how the non-local negativity interpolates between its exponentially small value at any non-zero $T$ to a non-zero, universal $O(1)$ value at $T = 0$. The scaling collapse of the data when plotted as a function of $e^{-L/\beta}$ again indicates that the quantum correlation length $\xi_Q \sim \beta$. We verified that the O(1) constant contribution to negativity at zero temperature agrees with the known result for a massless scalar \cite{krempa2017, Chen_2017}.

%
%
%Finally we comment on the eigenvalue structure of the matrix that determines the negativity of our model. We find that in the thermodynamic limit, the spectrum corresponding to our model in 1D consists of a `bulk' continuum part, and two isolated discrete  eigenvalues per entanglement cut which are isolated from the continuum spectrum  (see Fig.\ref{fig:ngvt_spectrum} and Appendix \ref{sec:eigenvalue} for detailed discussion). Interestingly, only the discrete eigenvalues less than unity contribute to the negativity, and the corresponding eigenfunction is localized at the entanglement cut, in contrast to the eigenfunctions of all other eigenvalues which possess plane-wave like features, and are spread out in the entire spatial region. This is indicative of the area-law for negativity of thermal states. We leave the full analytical analysis of the eigenvalue problem to the future. 
\begin{figure*}[ht]
	\centering
	\begin{subfigure}[b]{0.33\textwidth}
		\caption{}
		\includegraphics[width=\textwidth]{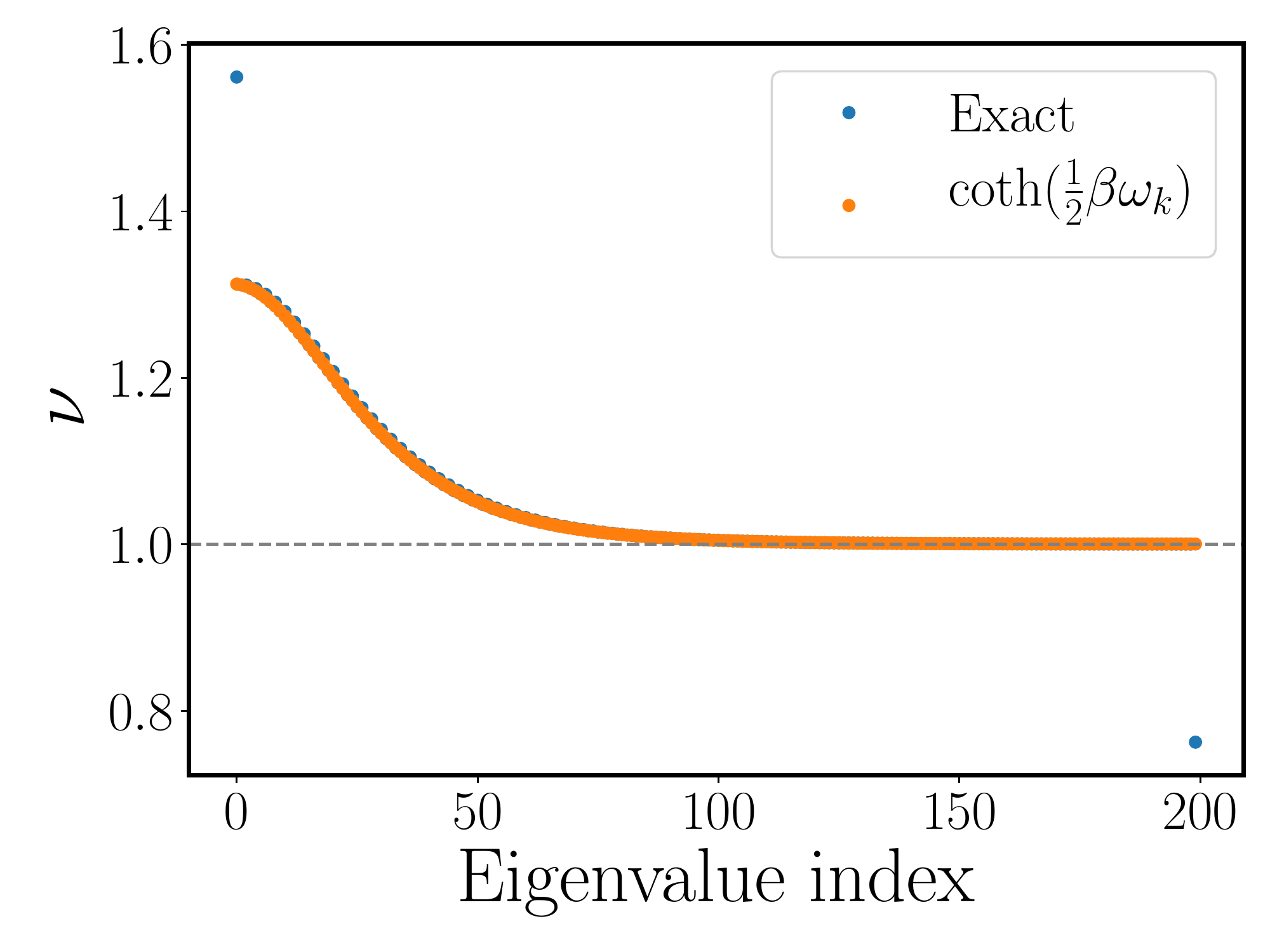}
		\label{fig:spectrum}
	\end{subfigure}
	\begin{subfigure}[b]{0.33\textwidth}
		\caption{}
		\includegraphics[width=\textwidth]{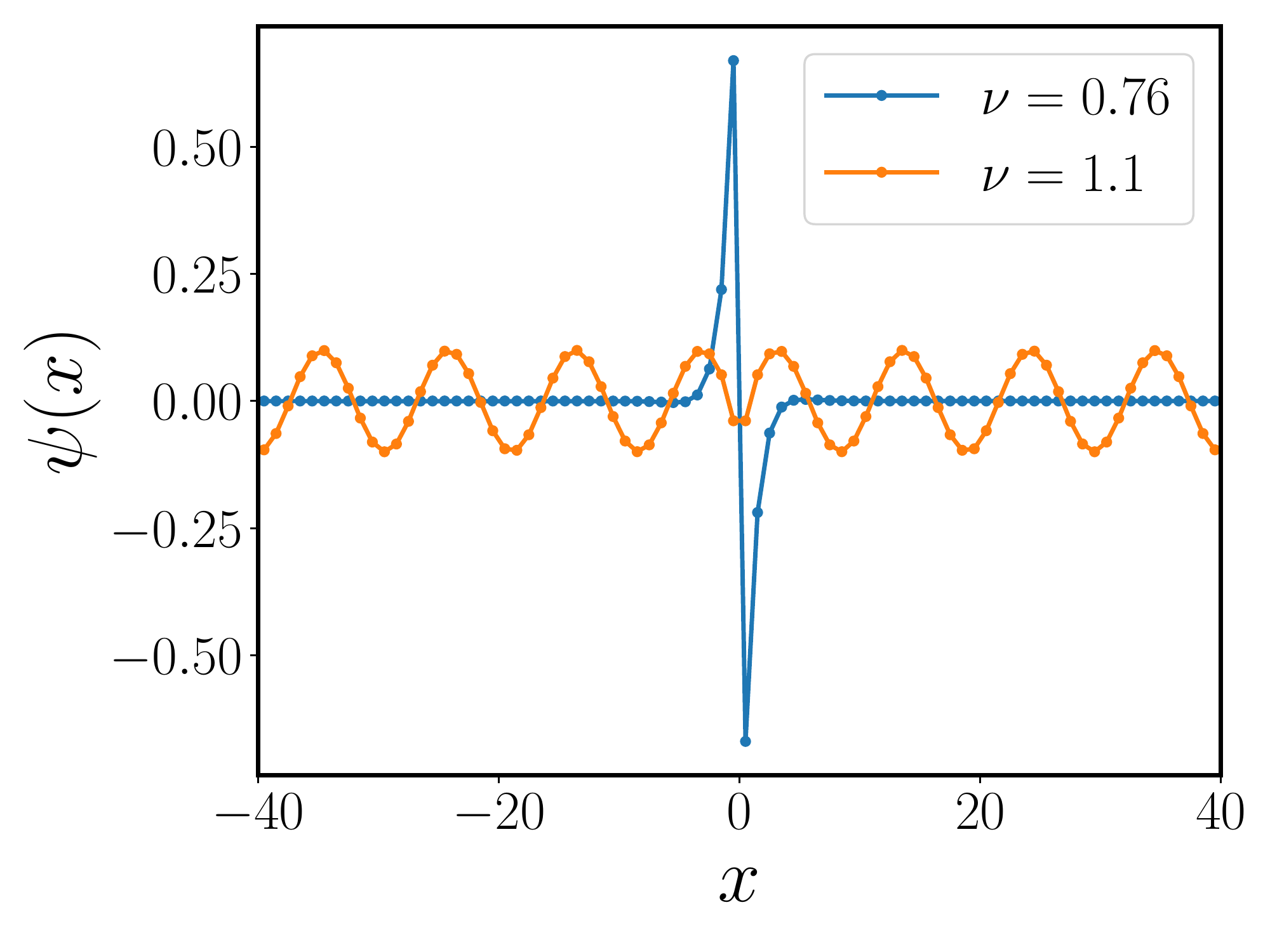}
		\label{fig:eigenfunction}
	\end{subfigure}
	\begin{subfigure}[b]{0.33\textwidth}
		\caption{}
		\includegraphics[width=\textwidth]{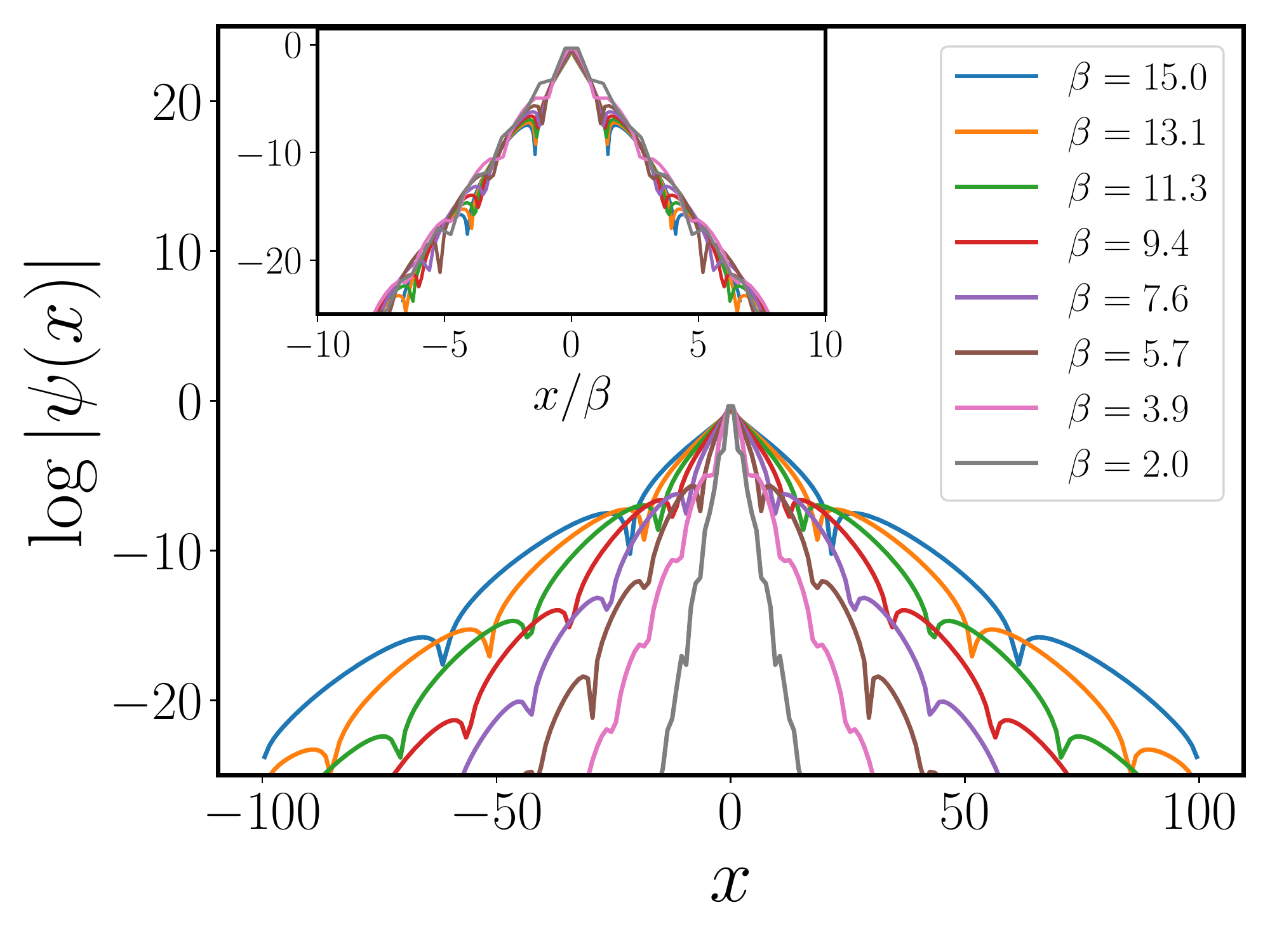}
		\label{fig:wf}
	\end{subfigure}
	\caption{Eigenvalue spectrum and eigenfunction for the matrix $\sqrt{\gamma_{\phi}P \gamma_{\pi}P}$ (see main text for details)  corresponding to the model defined  in Eq.\ref{eq:H} in one dimension with open boundary condition. We choose $L=200$ and $L_A=L/2$. (a) Eigenvalue spectrum of $\sqrt{\gamma_{\phi} P\gamma_{\pi}P} $ compared with $\coth(1/2\beta\omega_k)$, the eigenvalues of $\sqrt{\gamma_{\phi} \gamma_{\pi}}$ at $m =0.5$, and $T = 0.25$. (b) While the eigenfunction corresponding to $\nu=0.76$ is localized at the bipartition boundary ($x=0$), the eigenfunction corresponding to $\nu=1.1$, i.e. an eigenvalue in the bulk of the spectrum, behaves as a plane wave. (c) The eigenfunction with the lowest eigenvalue for $\sqrt{\gamma_{\phi} P \gamma_{\pi} P   }$ We choose $m=10^{-5}$ to simulate the massless limit. Inset: the same data plotted on a rescale horizontal coordinate $x/\beta$. } \label{fig:eigenspectrum}
\end{figure*}

\subsection{Eigenvalue structure of the partial transposed correlation matrix} \label{sec:eigenspectrum}

We find that the eigenvalues and eigenvectors of the correlation matrix that determines negativity have a very specific pattern which reveals more information about the mixed state entanglement. For a Gaussian density matrix such as ours, the negativity is determined by the eigenvalues of the matrix $\sqrt{\gamma_{\phi}P\gamma_{\pi}P}$ which we will denote as the `partial transposed correlation matrix'. Here $\gamma_{\phi}(\vec{r},\vec{r}')=2\expval{\phi_{\vec{r}}  \phi_{\vec{r}'}}$,  $\gamma_{\pi}(\vec{r},\vec{r}')=2\expval{\pi_{\vec{r}}  \pi_{\vec{r}'}}$, and 
$ P(\vec{r},\vec{r}') = \delta_{\vec{r},\vec{r}'} \quad \text{for}\quad  \vec{r} \in A$ and 
$-\delta_{\vec{r},\vec{r}'}   \quad \text{for}  \quad \vec{r} \in \bar{A}$. Specifically, $E_N=-\sum_{i=1}^{N} \min\{0,\log \nu_i\}$
where $\{\nu_i\}$ are the eigenvalues of $\sqrt{\gamma_{\phi}P\gamma_{\pi}P}$. Therefore only eigenvalues less than unity contribute to negativity.

 Consider, for instance, the  Hamiltonian in Eq.\ref{eq:H} in $d=1$ (as discussed above, for a bipartition without any corners, the eigenvalues in $d > 1$ can be determined in terms of eigenvalues for the $d=1$ problem). We find that the spectrum consists of a `bulk' continuum part and and two discrete eigenvalues isolated from the continuum spectrum  (see Fig.\ref{fig:spectrum}). We numerically find that the bulk continuum part in fact matches with the eigenvalues of $\sqrt{\gamma_{\phi} \gamma_{\pi}}$, the correlation matrix without any partial transpose operation. Quite strinkingly, as one notes from Fig.\ref{fig:spectrum}, only the isolated discrete eigenvalue less than unity contributes to the negativity.

Given the distinctive nature of eigenvalue spectrum, it is  instructive to contrast the  eigenfunctions corresponding to the continuum eigenvalues with that for the isolated eigenvalue that contributes to negativity. We find that while the eigenfunctions corresponding to the bulk continuum spectrum essentially behave as a plane wave, the eigenfunction corresponding to the discrete eigenvalue is localized at the bipartition boundary (see Fig.\ref{fig:eigenfunction}), signifying the fact that quantum entanglement is localized close to the boundary.  Furthermore, the localized eigenfunction decays exponentially away from the entanglement cut even in the massless limit (Fig.\ref{fig:wf}). In particular, the characteristic decay length is proportional to the inverse temperature $\beta$ as indicated by the scaling collapse analysis (see Fig.\ref{fig:wf} inset), quite similar to the aforementioned behavior of the `quantum correlation length' determined using the decay of the non-local negativity. 

%When the temperature $T$ is above the `sudden death temperature' $T_{sd}$, where negativity vanishes, the scaling collapse breaks down. This is consistent with the fact that for $T>T_{sd}$, entanglement cannot be captured by negativity. 

For plots in Fig.\ref{fig:eigenspectrum}, we impose  open boundary condition. We have checked that these observations apply to periodic boundary condition as well, the only difference being that now the discrete eigenvalues will be two-fold degenerate due to the presence of two entanglement cuts.

\section{Separability of bosonic Gaussian states with $U(1)$ and Time-reversal symmetries} \label{sec:bosonic_tb}

Above, we studied negativity for a Gaussian density matrix with Ising symmetry. In this section, we report a somewhat surprising observation on mixed state entanglement for a closely related problem. We consider a bosonic tight-binding model in arbitrary spatial dimension: 
\be 
H=-\sum_{ij}t_{ij} a_i^{\dagger} a_j, \label{eq:H_tightbinding}
\ee
where $i$ index labels the lattice sites and $a_i, a_{i}^{\dagger}$ are the corresponding annihilation and creation bosonic operator. Apart from the $U (1)$ symmetry corresponding to the particle number conservation, we also impose time-reversal symmetry so the hopping amplitude $t_{ij}$ is real. Crucially, we work in grand-canonical ensemble, i.e., we do not fix the number of bosons exactly, but only on average via a chemical potential (which will correspond to diagonal elements of the hopping matrix $t_{ij}$ in $H$). This model is of interest since it exhibits Bose-Einstein condensation (BEC) when cooled below a critical temperature. Therefore, one may wonder how quantum correlations behave across such a finite-$T$ phase transition. Surprisingly, we find that the Gibbs state, i.e. $\rho=\exp(-\beta H)/Z$, for this model can be written as a convex combination of product states and thus is \textit{separable}. Therefore, all measures of mixed state entanglement, including negativity, are identically zero at all temperatures. It is important to note that negativity can be zero even for non-separable states \cite{Horodecki_bound, horodecki_revmodphys}, and therefore this result is much stronger than just showing that negativity is zero for this system.

 The central idea in our proof is to employ the  Glauber-Sudarshan `$P$ representation' \cite{Sudarshan,Glauber} for the density matrix $\rho$:

\begin{equation}\label{eq:rho_decompose}
\rho=\int_{\mathbb{C^{N}}} \prod_{i=1}^N d^2\alpha_i    P(\bm{\alpha } )    \ket{\bm{\alpha}}\bra{\bm{\alpha}},
\end{equation}
where $\ket{\bm{\alpha}}=\otimes_i \ket{\alpha_i}$ is a tensor product of coherent states at all sites. $P(\bm{\alpha } ) $ is a quasiprobability distribution since it can be negative in general. We find that when $\rho$ is a Gibbs state corresponding to the aforementioned tight-binding model, $P(\bm{\alpha } ) $ is a proper probability distribution function, and thus $\rho$ is separable for all inverse temperature $\beta$. See appendix \ref{sec:bosonic_separability} for details.

At a first glance, this result seems puzzling since as $T\to 0 $, one might expect that $\rho$ will correspond to a pure ground state of $H$, which can be entangled, and is contrary to our finding. This tension is resolved by noticing that $\rho$ is not pure even at $T=0$. To see this, consider the thermal state $\rho \propto e^{-\beta H}$ for a single mode Hamiltonian $H=\epsilon a^{\dagger}a$. A simple calculation shows purity $\tr \rho^2=1/(2\expval{a^{\dagger}a  }_\beta+1)$, where $\expval{a^{\dagger}a  }_\beta$ is the expectation value of $a^{\dagger}a$ in the thermal state, and thus the state is never pure for any non-zero number of bosons. Alternatively, the tension can be traced back to the difference between the canonical ensemble and the grand canonical ensemble. As mentioned above, the Gibbs state $\rho \propto \exp{-\beta H}$ is treated within grand canonical ensemble, where the particle number is fixed only on average by a chemical potential. On the other hand, in the canonical ensemble, the Gibbs state is restricted to a fixed particle number sector $\rho\propto \exp{-\beta H} \delta(N- \sum_i a_i^{\dagger }a_i )$. Due to the delta function constraint, all the bosons are enforced to occupy the lowest single particle state, which is a pure state that may very well be entangled.

\section{Summary and Discussion} \label{sec:discuss}

In this work we set out to reconcile the tension between the following two observations: (a) The universal long-distance correlations for typical finite-$T$ transitions are described by a low-energy effective theory that is fully classical. \cite{sachdev2011quantum} (b) The mixed state quantum entanglement, as quantified by entanglement negativity, is singular across several such transitions \cite{lu2018singularity}. We studied a specific model that hosts a finite-$T$ transition in a mean-field universality class to understand and resolve this tension. Conceptually, our basic idea is to separate the negativity into a local term, i.e., a term which can be written as sum of local terms along the entangling boundary, and contributes to the leading area-law behavior, and a non-local term, which cannot be written in this way and therefore encodes long-distance quantum correlations. We found that in the thermodynamic limit, the singularity of negativity originates only from the local term, and can be fully canceled out by a subtraction scheme that leaves only the non-local term, which vanishes exponentially with the total system size $L$. Therefore, in the model studied, the long-distance quantum correlations are non-singular across the transition in the thermodynamic limit. We defined a length scale $\xi_Q$ over which quantum correlations exist, and showed that at non-zero $T$, $\xi_Q$ remains finite even when the physical correlation length $\xi$ diverges. Therefore quantum mechanically, the system continues to be short-range correlated, despite a diverging physical correlation length. This provides a sharp distinction between a `quantum phase transition' and a `classical phase transition': the quantum correlation length diverges only at a quantum phase transition.

Our discussion was focused on a Gaussian theory that exhibits mean-field critical exponents. Interactions at a finite $T$-transition in a quantum Ising model modify the critical exponents, but the critical field theory again belongs to the  universality class of a classical Ising model \cite{sachdev2011quantum}. Therefore, our expectation is that the tripartite negativity $\Delta_3 E_N$ will continue to decay exponentially with system size even at the Wilson-Fisher fixed point. Note that it is already rather non-trivial to find a quantity, namely $\Delta_3 E_N$, that decays exponentially in the Gaussian  critical theory. All correlation functions of local operators decay as power-law, simply because the physical correlation length is infinite. As an analogy, consider the quantum phase transition in the quantum Ising model at $T = 0$. Quantum entanglement at the interacting fixed-point has the same general structure as that at the Gaussian critical theory at $T = 0$, the only difference being that the value of the universal $O(1)$ subleading term in entanglement is modified \cite{metlitski2009,whitsitt2017entanglement,Inglis_2013,casini07}. It will of course be very interesting to do an actual calculation of negativity at the finite-$T$ transition using field-theoretic techniques, to check this intuition.

We also studied the singularity associated with the area law coefficient in detail. Although all our calculations are restricted to the Hamiltonian in Eq.\ref{eq:H}, we expect these conclusions to generalize to other finite-$T$ transitions. Based on results in Ref.\cite{lu2018singularity}, we conjecture that the leading singular part of area-law coefficient originates from the expectation value of a local operator that is invariant under the symmetries of the Hamiltonian, and has the lowest scaling dimension. In the case of Ising model, this operator corresponds to the energy density, and therefore, we expect

\be 
E_{N,\textrm{local, singular}} \sim |t|^{1-\alpha} L^{d-1} \label{eq:singular_ngvt}
\ee 
where $t$ is the thermal tuning parameter i.e. $t = (T-T_c)/T_c$ or $t = g-g_c$, and $\alpha$ is the critical exponent that defines the divergence of specific heat $C \sim |t|^{-\alpha}$ (and consequently, the energy density has a singular part that scales as $|t|^{1-\alpha}$). This scaling matches with the results for our mean-field model: within mean-field theory, $\alpha = 0$, and therefore $E_{N,\textrm{local, singular}} \sim |t| \sim m^2$.

As shown in Ref.\cite{vidal2002}, negativity upper bounds `entanglement of distillation', which intuitively corresponds to best rate at which one can extract near-perfect EPR singlets from multiple copies of a state using local operations and classical communications (LOCC). The absence of long-range negativity in our model even at the finite-$T$ critical point indicates that the distilled EPR pairs originate only from correlations close to the entangling boundary, as suggested in Fig.\ref{fig:lengths}. Relatedly, one expects that operators that contribute to the violation of Bell's inequality between regions $A$ and $\bar{A}$ are located close to the entangling boundary.

Motivated by the calculation of negativity in our model that has a $\mathbb{Z}_2$ symmetry, we also studied negativity in closely related models that instead have an $U(1)$ symmetry.  We found a surprising result that the thermal density matrices corresponding to bosonic tight-binding models with $U(1)$ and time-reversal symmetry are separable, and therefore, any mixed state measure of entanglement, including negativity, vanishes for such states. One consequence of this result is that a convex sum of density matrices $\rho = \sum_i p_i \rho_i$, where each $\rho_i$ is Gaussian and has $U(1)$ and time-reversal symmetry, is also separable. Note that $\rho$ itself is not Gaussian and corresponds to an interacting, albeit generically non-local, Hamiltonian.

Our results raise an intriguing question. It has been argued that a thermal state can be efficiently prepared if a system does not possess finite-$T$ topological order, and if it is above any finite-$T$ phase transition, so that the correlation functions of local operators are short-ranged \cite{swingle2016_1, swingle2016_2, Brandao2019, martyn2018, wu2018variational,Cottrell2019, maldacena2018eternal,chapman2019complexity}. Our results indicate that quantum correlations can be short-ranged even when the correlation functions of local operators are long-ranged. Although our calculations were specific to a rather simple Hamiltonian (Eq.\ref{eq:H}), we suspect that this is true more generally as long as finite-$T$ topological order is absent \cite{hastings2011}. This raises the possibility that even density matrices that have infinite correlation length, such as quantum systems below or even at a symmetry breaking finite temperature transition, might be efficiently preparable.  A starting point could be to consider a purely classical density matrix below or at the critical temperature $T_c$, and variationally apply a finite-depth quantum channel on it, so as to minimize the trace distance between the resulting density matrix and the actual density matrix $\rho \propto e^{-\beta H}$.

It's worth comparing our protocol with other measures introduced previously to detect quantum coherence at finite temperature. In particular, Ref.\cite{malpetti_qcf} introduced a measure called `quantum correlation function' (QCF) that takes the form $\langle \delta O_A \, \delta O_B \rangle_{Q} =  \langle \delta O_A \, \delta O_B \rangle - \frac{1}{\beta}\int_{0}^{\beta} d\tau \langle \delta O_A(\tau)\, \delta O_B(0) \rangle$ where $\delta O = O - \langle O\rangle$, and all averages are with respect to the thermal density matrix. It was argued in Ref.\cite{malpetti_qcf} that this quantity is smooth across finite-$T$ transitions, and decays exponentially, allowing one to define a `quantum coherence length'. One advantage of QCF is that it is relatively simple to calculate compared to entanglement based measures such as negativity. On the other hand, due to its definition in terms of local operators, QCF is not suitable to capture non-local \textit{many-body} entanglement. As a concrete illustration, consider a gapped, topologically ordered phase at zero temperature. Since $\Delta_3 E_N$ equals the topological entanglement entropy, the `quantum correlation length' introduced in our paper is infinite throughout the gapped phase. In strong contrast, the `quantum coherence length' based on QCF is finite and just equals the correlation length defined via local operators.

%An application was discussed in Ref.\cite{Shapourian2019}, where it was shown that for the thermal state of fermions at finite density, one obtains an area-law negativity in contrast to $L_A^{d-1}\log(L_A)$ scaling seen at $T =0$ \cite{Gioev06, wolf06}.

Let us mention a few other future directions. Since our proposed method is designed to isolate the non-local part of negativity, it will be worthwhile to apply it to characterize finite-$T$ topological order in models such as 4D Toric code \cite{dennis2002, hastings2011}. It will also be  interesting to study Renyi versions of negativity for interacting models using quantum Monte Carlo method \cite{chiamin2014, kallin2010, Isakov12,yingjer2019}, and implement the subtraction scheme introduced here numerically. Another  direction is to study negativity for interacting fermion systems that exhibit finite temperature phase transition such as the model of two SYK systems \cite{sachdev1993, kitaev_kitp} coupled to each other \cite{maldacena2018eternal}. We note that for fermions, two different definitions of partial transpose have been introduced \cite{Eisler_2015, Shapourian2017,Shapourian2018} and it will be interesting to understand the qualitative differences between the two in interacting theories.

\begin{acknowledgments}
	\emph{Acknowledgments:} 
	TG is supported by an Alfred P. Sloan Research Fellowship. We thank Tim Hsieh and Sung-Sik Lee for helpful discussions, and especially John McGreevy for helpful discussions and comments on the manuscript. We thank Perimeter Institute for Theoretical Physics for their hospitality where part of this manuscript was completed.
\end{acknowledgments}

\bibliography{v1bib}
\renewcommand\refname{Reference}
% \bibliographystyle{unsrt}

%\bibliography{bib_appendix.bib}
%\renewcommand\refname{Reference}
%% \bibliographystyle{unsrt}
\newpage

\appendix

\section{von Neumann entropy in the Gaussian theory} \label{sec:entropy_gaussian}

Consider a $d$ dimensional system of size $L$ described by the classical Hamiltonian
\begin{equation}
H=\frac{1}{2}\int d^dx~m^2 \phi(x) + \left(\nabla\phi(x) \right)^2,
\end{equation}
we prove that as $mL \to 0$, $L\to \infty$, the von Neumann entropy of the Gibbs state $\rho\sim \exp{-\beta H}$ contains a subleading term $\Delta S= -\log mL$ with the periodic boundary condition imposed on all spatial directions.

To proceed, we first discretize the continuum theory in a finite box of volume $V=L^d$ with lattice cutoff being $1$. A standard calculation of the Gaussian theory gives the thermal partition function 
\begin{equation}
Z=\sqrt{\frac{   \left(2\pi T\right)^V}{ \prod_{k} \lambda_k  }}, 
\end{equation}
where $\lambda_k=m^2+4\sum_{n=1}^d \sin^2(\frac{1}{2} k_{i}  )$, and $k_i\in 2\pi n/L$ with $n=0,\cdots,L-1$. It follows that the thermal free energy $F$ is
\begin{equation}
F=-T\log Z= -\frac{1}{2}VT \log (2\pi T)+   \frac{1}{2} T	\log \left[\prod_{k} \lambda_k \right],
\end{equation}
and the von Neumann entropy is given by $S=-\partial F/\partial T$:
\begin{equation}\label{appendix:entropy}
S=\frac{V}{2}\left[ 1+\log \left(   2\pi T \right)  \right]  - \frac{1}{2}  \log \left[\prod_{k} \lambda_k \right].
\end{equation}
Since the first term contributes to the volume law part of $S$, we will only focus on the second term. To proceed, we employ the identity\cite{Brankov_1993}
\begin{equation}
\prod_{n=0}^{L-1} \left[ a^2+4\sin^2( n\frac{\pi }{L})      \right]  =g(a),
\end{equation}
where $g(a)= \left\{  2^{-L} \left[ \sqrt{a^2+4}+a   \right]^L -2^{L} \left[ \sqrt{a^2+4}+a   \right]^{-L}   \right\}^2$. Thus one can first evaluate the product along the $d$'th spatial direction to obtain 
\begin{equation}
\prod_k\lambda_k=    \prod_{ \substack{   {k_1,\cdots,k_d}  }} \lambda_k=\prod_{ \substack{   {k_1,\cdots,k_{d-1}}   }} g\left(\sqrt{m^2+4\sum_{n=1}^{d-1} \sin^2(\frac{1}{2} k_{i}  )}\right).  
\end{equation}
By singling out the contribution of the zero mode, one finds 
\begin{equation}
\prod_k\lambda_k= g(m) \prod_{ {(  k_1,\cdots,k_{d-1})    \neq 0 }   } g\left(\sqrt{m^2+4\sum_{n=1}^{d-1} \sin^2(\frac{1}{2} k_{i}  )}\right).  
\end{equation}
As $L\to \infty$ with $mL\to 0$, one finds $g(m) \to (mL)^2$, and thus 
\begin{equation}
\log \left[ \prod_k \lambda_k  \right] = 2\log (mL) + \sum_{(  k_1,\cdots,k_{d-1})    \neq 0    }  \log \left[ g\left( 2\tilde{\omega} \right)  \right],
\end{equation}
where $\tilde{\omega}=\sqrt{\sum_{n=1}^{d-1} \sin^2(\frac{1}{2}k_i)}$.
By noticing that as $\tilde{\omega}=O(1)$, $L\to \infty$, $\log g(2\tilde{\omega})  =  2L   \log \left[   \left(\sqrt{\tilde{\omega}^2+1}+\tilde{\omega}\right)  \right]$. We find 
\begin{equation}
\log \left[ \prod_k \lambda_k  \right] \to 2\log (mL)  + 2L^d \int  \frac{d^{d-1}k}{(2\pi)^{d-1}}  \log \left[   \sqrt{  \tilde{\omega}^2  +1}  +\tilde{\omega}   \right].
\end{equation}
Note that the above result is not exact since $\tilde{\omega}$ can be of order $1/L$, in which case $\log g(2\tilde{\omega})\sim O(1)$ instead of $O(L)$. Nevertheless it does not affect the logarithmic divergence caused by the first term. By plugging this result into Eq.\ref{appendix:entropy}, we find the logarithmic divergence for the entropy $S\sim -\log (mL)$ for $mL\to 0$ at arbitrary dimension. In this calculation, we start from a purely classical Gaussian theory for deriving the logarithmic divergence in the massless limit. However, this result is applicable to the quantum Gaussian model studied in the main text as well by mapping a $d$ dimensional quantum problem to a classical problem with one extra dimension of size $\beta$.

\section{Mapping of negativity of a d-dimensional problem to a 1-dimensional problem}  \label{sec:mapping1d}

Consider a $d$ dimensional lattice of $N$ sites with the Hamiltonian
\begin{equation}
H=\frac{1}{2} \sum_{\vec{r}} \left( \pi_{\vec{r}}^2 +m^2 \phi_{\vec{r}}^2 \right) +  \frac{1}{2}  \sum_{\expval{\vec{r},\vec{r}'}}K \left( \phi_{\vec{r}} -\phi_{\vec{r}'}   \right) ^2,
\end{equation}
Imposing the periodic boundary condition for all spatial direction, a standard calculation for two-point functions gives  
\begin{equation}
\begin{split}
&  \gamma_{\phi}(\vec{r},\vec{r}')=2\expval{\phi_{\vec{r}}  \phi_{\vec{r}'}}=\frac{1}{N} \sum_{\vec{k}} e^{i\vec{k}\cdot \left(\vec{r} -\vec{r}'  \right)} \frac{1}{\omega_{\vec{k}}} \coth(\frac{1}{2} \beta \omega_{\vec{k}}) \\
&  \gamma_{\pi}(\vec{r},\vec{r}')=2\expval{\pi_{\vec{r}}  \pi_{\vec{r}'}}=\frac{1}{N} \sum_{\vec{k}} e^{i\vec{k}\cdot \left(\vec{r} -\vec{r}'  \right)} \omega_{\vec{k}} \coth(\frac{1}{2} \beta \omega_{\vec{k}}) ,
\end{split}
\end{equation}
where $\vec{k} =\left(k_1,k_2,\cdots,k_d   \right)=\frac{2\pi}{L} \left(n_1,n_2,\cdots ,n_d\right)$ for $n_i=0,1,\cdots L-1$, and  $\omega_{\vec{k}}= \sqrt{m^2+4K\sum_{i=1}^d \sin^ 2(\frac{1}{2} k_{i}  )    }$. To calculate the negativity between the subregion $A$ and its complement $B$, we follow the correlation matrix technique introduced in Ref.\cite{audenaert2002entanglement}. One defines a matrix $P$ with the matrix element 
\begin{equation}
P(\vec{r},\vec{r}') =\begin{cases}
\delta_{\vec{r},\vec{r}'}   ~~ \quad \text{for}\quad  \vec{r} \in A\\
-\delta_{\vec{r},\vec{r}'}   \quad \text{for} \quad \vec{r} \in B,
\end{cases}
\end{equation}
and the negativity is given by 
\begin{equation}\label{append:eq:negativity}
E_N=-\frac{1}{2}\sum_{i=1}^{N} \min\{0,\log \lambda_i\},
\end{equation}
where $\lambda_i$ is the eigenvalue of $\gamma_{\phi}P\gamma_{\pi}P$.

Suppose that the boundary between $A$ and $B$ only cut through one of the spatial direction, say direction labeled by $1$, the matrix $\gamma_{\phi}P \gamma_{\pi} P$ preserves the translational invariance along the other $d-1$ directions, and therefore its eigenfunction $\psi(\vec{r})$ takes the form 
\begin{equation}
\psi(\vec{r})=e^{i\sum_{i=2}^d k_i r_i   } \psi_1(r_1).
\end{equation}
Given this ansatz, one can reduce the $d$-dimensional problem to a $1$-dimensional problem. In particular, the negativity is
\begin{equation}
E_N=\sum_{k_2,\cdots k_d} E_N^{(1)}(k_2,\cdots,k_d),
\end{equation}
where $ E_N^{(1)}(k_2,\cdots,k_d)$ is the negativity of the $1$-dimensional theory with a modified mass term :
\begin{equation}
m^2\to m^2+ 4K\sum_{i=2}^d\sin[2](\frac{1}{2}k_i ).
\end{equation}
Note that this implies the leading contribution in $E_N$ is an area law term if $E_N^{(	1)}$ in $1$-dimension follows an area law.

\section{Negativity between two sites}\label{sec:two_sites}
Consider a system of two sites, the dispersion relation reads $\omega_k=\sqrt{m^2+4\sin^2( k/2)  }$ with $k=0,\pi$, and the covariance matrices are $\gamma_{\phi}=\begin{pmatrix} a_{+}& a_{-}\\ a_{-}&a_{+}  \end{pmatrix}$, $\gamma_{\pi}=\begin{pmatrix} b_{+}& b_-\\ b_-&b_{+}  \end{pmatrix}$ where 
\begin{equation}
\begin{split}
&a_{\pm }=\frac{1}{2} \left[  \frac{1}{m} \coth(\frac{1}{2}\beta m )   \pm \frac{1}{\sqrt{m^2+\frac{1}{4}}}\coth(\frac{1}{2} \beta \sqrt{m^2+4}  )  \right]\\
&b_{\pm}=\frac{1}{2} \left[  m \coth(\frac{1}{2}\beta m ) \pm   \sqrt{m^2+\frac{1}{4}}\coth(\frac{1}{2} \beta \sqrt{m^2+4}  )  \right].
\end{split}
\end{equation}
The eigenvalues of $\gamma_{\phi}P  \gamma_{\pi}  P$ are

\begin{equation}
\begin{split}
&\lambda_1= \frac{\sqrt{m^2+4}}{m} \coth(\frac{1}{2}  \beta m  )  \coth(\frac{1}{2  } \beta \sqrt{m^2+4}  )\\
&\lambda_2=\frac{m  }{ \sqrt{m^2+4}  }   \coth(\frac{1}{2}  \beta m  )  \coth(\frac{1}{2  } \beta \sqrt{m^2+4}  ).
\end{split}
\end{equation}

%
%\begin{equation}
%\begin{split}
%&\lambda_1=(a_{+} +a_{-}  ) (b_{+} -b_-  )= \frac{\sqrt{m^2+4}}{m} \coth(\frac{1}{2}  \beta m  )  \coth(\frac{1}{2  } \beta \sqrt{m^2+4}  )\\
%&\lambda_2=(a_{+} -a_{-}  ) (b_{+} +b_-  )  =\frac{m  }{ \sqrt{m^2+4}  }   \coth(\frac{1}{2}  \beta m  )  \coth(\frac{1}{2  } \beta \sqrt{m^2+4}  ).
%\end{split}
%\end{equation}

While $\lambda_1\geq 1$ for arbitrary $\beta$ and $m$, $\lambda_2$ can be less than $1$ for some parameters. Thus the negativity is $E_N =  -\frac{1}{2}  \min\{0,\log \lambda_2\}$. In particular, while $\lambda_1$ diverges at the massless limit $m\to 0$, the eigenvalue $\lambda_2$, which contributes to the negativity, remains finite and is perturbative in $m^2$. Specifically, $-\log \lambda_2= -\log (\frac{\coth \beta }{\beta})   +  \left( \frac{1}{8}- \frac{\beta^2}{12}   +\frac{\beta}{  8} \coth\beta-\frac{\beta}{8}\tanh\beta  \right)m^2+O(m^4)$. Thus the leading order correction for the negativity in the massless limit is of order $O(m^2)$.

\section{Separability of the thermal density matrix corresponding to the bosonic tight-binding model} \label{sec:bosonic_separability}

We define the $N$-sites coherent state: $\ket{\bm{\alpha}}=\ket{ \alpha_1,\cdots, \alpha_N }$ such that $a_{i} \ket{\bm{\alpha}}= \alpha_i  \ket{\bm{\alpha}}$. Note that the coherent state $\ket{\bm{\alpha}}$ can be generated by acting the displacement operator $\hat{D}(\bm{\alpha })\equiv  \exp{\sum_{i=1}^N \left[  \alpha_ia_i^{\dagger}  -\alpha_i^{*}  a_i \right]  }$ on a vacuum state: $\ket{\bm{\alpha}}= D(\bm{\alpha }) \ket{0}$. 

The central ingredient we employ is the Glauber-Sudarshan P representation\cite{Sudarshan,Glauber},  which can cast any bosonic state into a diagonal matrix in the basis of coherent states:
\begin{equation}\label{eq:rho_decompose2}
\rho=\int_{\mathbb{C^{N}}} \prod_{i=1}^N d^2\alpha_i    P(\bm{\alpha } )    \ket{\bm{\alpha}}\bra{\bm{\alpha}},
\end{equation}
where $d^2\alpha_i \equiv d\text{Re}(\alpha_i) d\text{Im}(\alpha_i)$. To define $P(\bm{\alpha})$, we first introduce the characteristic function $\chi(\bm{\alpha})$\cite{serafini2017quantum}: 
\begin{equation}
\chi(\bm{\alpha})= \tr \left[ \hat{D}( \bm{\alpha}  )\rho    \right]  e^{\frac{1}{2}   \sum_{i=1}^N  \abs{\alpha_i}^2}.
\end{equation} 
$\chi(\alpha)$ serves as a generating function in the sense that its derivatives give the expectation value of the observables: 

\begin{equation}
\tr\left[ \left(a_i^{\dagger}\right)^m a^n_j  \rho \right] = \left. \left[  \left(   \frac{\partial }{\partial \alpha_i}   \right)^m \left(  - \frac{\partial }{\partial \alpha^{*}_j  }   \right)^n  \chi(\bm{\alpha}) \right] \right|_{\bm{\alpha}=0}.
\end{equation}
Then $P(\bm{\alpha})$ is given by the complex Fourier transform of $\chi$:
\begin{equation}
P(\bm{\alpha})=\frac{1}{\pi^{2N}}\int_{\mathbb{C^{N}}} \prod_{i=1}^N d^2\gamma_i e^{\sum_{i=1}^N   \alpha_i \gamma_i^*-\alpha_i^*\gamma_i     }   \chi(\bm{\gamma}).  
\end{equation}

Up until this point, the discussion applies to any bosonic state and generically, $P(\bm{\alpha } )$ is not necessarily positive. Now we specialize to Gaussian states, such as the tight-binding model of bosons (Eq.\ref{eq:H_tightbinding}), the subject of our focus. By defining $\alpha_i =\frac{1}{\sqrt{2}}  \left( x_i +ip_i  \right)$, one can show that for a Gaussian state, $P$ can be expressed as \cite{serafini2017quantum}:

\begin{equation}
P(\mathbf{r})=\frac{1}{\left( 2\pi^2 \right)^N}   \int_{\mathbb{R}^{2N}} d\mathbf{r'} e^{i\mathbf{r}'^T  \left( \Omega \mathbf{r}   + \Omega^T  \overline{\mathbf{r}}   \right)  }     e^{\frac{1}{4}    \mathbf{r}'^T   \left(  1-\Omega^T \sigma  \Omega  \right) \mathbf{r}'  },
\end{equation}
Here $\mathbf{r}^T=(x_1,p_1,\cdots,x_N,p_N)$ with $\overline{\mathbf{r}} $ being the first moment ($\overline{r}_i= \tr(\rho r_i)$) and $\sigma$ being the $2N\cross 2N$ covariance matrix ($\sigma_{ij}=\tr(\rho \{r_i,r_j \})$). $\Omega$ denotes the symplectic matrix: $ \Omega= \bigoplus_{i=1}^N \begin{pmatrix} 0 &1\\-1&0  \end{pmatrix}$. The state $\rho$ will be separable if  $ 1-\Omega^T \sigma  \Omega \leq 0$ since then $P(\bm{r})$ is a well-defined probability function. Below we show that a bosonic tight-binding model with $U(1)$ and time-reversal symmetry (Eq.\ref{eq:H_tightbinding}) indeed satisfies this condition, which proves the separability in the form of Eq.\ref{eq:rho_decompose2}. 

\subsection{Proof that $\Omega^T\sigma \Omega \geq 1$}

We divide the proof into two-parts, we first show that $\sigma \geq  1$, i.e., all eigenvalues of $\sigma$ are greater than unity, and next we show that $\Omega^T\sigma \Omega$ and $\sigma$ have identical eigenspectrum, and thus $\Omega^T\sigma \Omega \geq 1$. 

\subsubsection{Proof that $\sigma \geq 1$}

Given $H=-\sum_{i,j} t_{ij}a^{\dagger}_{i}a_{j}$ on a $N$-site lattice in arbitrary spatial dimension, we can define the conjugate variables:  $x_i=\frac{1}{\sqrt{2}}\left(a_i+a_{i}^{\dagger}    \right)$ and $p_i=\frac{1}{i\sqrt{2}}\left(a_i-a_{i}^{\dagger}    \right)$. By introducing $\bm{r}^T=\left( x_1,\cdots, x_N, p_1,\cdots ,p_N   \right)$, our goal is to calculate the covariance matrix  $\sigma_{ij}  =\expval{\{r_i-\overline{r}_i , r_j -\overline{r}_j    \}   }  $, where $\{A,B\}=AB+BA$, and $\overline{r_i}=\expval{r_i}$ with the expectation values taken with respect to a thermal state $\rho=e^{-\beta H}/Z$. To proceed, we write $a_i=\sum_{k} u_{ik} b_k$ to diagonalize the Hamiltonian:  $H=\sum_kE_k b_{k}^{\dagger} b_k$. Given this result, we find
\begin{equation}
\begin{split}
&\expval{ \{ x_i,x_j \}  }=  \expval{ \{ p_i,p_j \}  }=\delta_{ij}+  2\sum_{k} \text{Re}\left[  u^*_{ik} u_{jk}  \right]  \expval{b_{k}^{\dagger} b_k } \\
&\expval{ \{ x_i,p_j \}  }  =2\sum_{k}  \text{Im}\left[u_{ik}^{*}u_{jk}  \right] \expval{b_{k}^{\dagger} b_k },
\end{split}
\end{equation}
and $\overline{r_i}=0$. Note that $u$ can be chosen as an orthogonal matrix since $t_{ij}$ is real (due to time reversal symmetry of $H$). Therefore, $\sigma=\gamma_x \bigoplus \gamma_p$, with 
\begin{equation}
\gamma_{x,ij}= \gamma_{p,ij}=  \delta_{ij}+2\sum_{k}  u_{ik} \expval{b_{k}^{\dagger} b_k } u_{jk},
\end{equation}
or equivalently, $\gamma_x=u\Lambda u^{\dagger}$, where $\Lambda=\text{ diag } \left\{  1+2\expval{b_k^{\dagger}b_k  }   \big\vert \text{for }  k=1,2,\cdots,N \right\}  $ storing the eigenvalues of $\gamma_x$. This completes the proof that $\sigma =\gamma_x \bigoplus \gamma_p \geq 1$.

\subsubsection{Proof that $\sigma$ and $\Omega^T\sigma \Omega$ have identical spectrum}

 To prove this, first notice that an orthogonal matrix $O$ can diagonalize $\sigma$ since it is symmetric: $\sigma=u\Lambda u^T$, where $\Lambda$ is a diagonal matrix. Then,
\begin{equation}
\Omega^T\sigma \Omega  = \Omega^T O\Lambda O^T  \Omega.
\end{equation}
Recall that $\Omega^T\Omega =\Omega \Omega^T = -1$, we then have 
\begin{equation}
\Omega^T\sigma \Omega  = \left(\Omega^T O \Omega \right) \left( \Omega^T \Lambda \Omega \right)    \left(  \Omega^T   O^T  \Omega  \right) = O' \Lambda' O'^T  ,
\end{equation}
where $O'\equiv \Omega^T O \Omega $ and $\Lambda'  = \Omega^T \Lambda \Omega$. A straightforward calculation shows that $O'$ is an orthogonal matrix, and $\Lambda'$ is a diagonal matrix. In particular, the set of the diagonal entries of $\Lambda'$ is exactly the same as that of $\Lambda$, and hence  $\Omega^T\sigma \Omega $ and $\sigma$ have the same eigenspectrum. Explicitly, if 
\begin{equation}
\Lambda=\text{diag}\left( a_1,b_1,a_2,b_2,\cdots    \right),
\end{equation}
then
\begin{equation}
\Lambda'=\text{diag}\left( b_1,a_1,b_2,a_2,\cdots    \right).
\end{equation}

\subsection{An aside: alternative proof for zero negativity for bosonic tight-binding model}

Recall that for a separable state, all measures of entanglement, including negativity, are identically zero, but a state with zero negativity can still be non-separable \cite{Horodecki_bound, horodecki_revmodphys}. Therefore, the above proof already establishes that negativity corresponding to a tight-binding Hamiltonian of bosons with $U(1)$ and time-reversal is identically zero. Still, as an aside, we provide an alternate proof of zero negativity for this system.

 From the fact that the eigenvalues of the correlation matrix $\gamma_x$ are greater than or equal to $1$, i.e. $\gamma_x\geq 1$, we prove that $\gamma_xP\geq 1$, and thus all eigenvalues of $\gamma_xP\gamma_xP$ are greater than or equal to 1. It follows that the negativity is zero from Eq.\ref{append:eq:negativity}. 

\noindent\underline{Proof for $\gamma_xP \geq  1$}: \\
Given a $N \cross N$ symmetric matrix $\gamma_x$ with all eigenvalues $\lambda_m \geq 1$ and a diagonal matrix $P$ with $1$ or $-1$ diagonal entries. First we perform an eigendecomposition of $\gamma_x$: 
\begin{equation}\label{Append:eigen_x}
\gamma_x=O^T\Lambda O, 
\end{equation} 
where $O$ is an orthogonal matrix and $\Lambda=\text{diag} \left(\lambda_1, \cdots,  \lambda_N  \right)$. Next the eigenequation of $\gamma_xP$ reads 
\begin{equation}
\gamma_xP v_m =\tilde{\lambda}_m v_m,
\end{equation}
where $\tilde{\lambda}_m$ and $v_m$ are the eigenvalue and eigenvector respectively. Taking the transpose of the above equation gives 
\begin{equation}
v_m^T P\gamma_x    =\tilde{\lambda}_m v^{T}_m.
\end{equation}
From the product of the above two equations and employing Eq.\ref{Append:eigen_x}, we obtain 
\begin{equation}
\tilde{\lambda}^2_m=\tilde{v}_m^T\Lambda\tilde{v}_m=\sum_{i=1}^N  \tilde{v}^2_{m,i} \lambda_i^2,
\end{equation}
where $\tilde{v}_m=OPv_m$. Since $\tilde{v}_m^{T}\tilde{v}_m=\sum_{i=1}^N \tilde{v}^2_{m,i} =1$, $\tilde{\lambda}^2_m\geq \sum_{i=1}^N  \tilde{v}^2_{m,i} \lambda_{\text{min}}^2 \geq 1$, where $\lambda_{\text{min}}$ is the minimum eigenvalue of $\gamma_x$. Thus, all eigenvalues $\tilde{\lambda}_m$ of $\gamma_xP$ satisfy $\tilde{\lambda}_m \geq 1$.

 \end{document}